\documentclass[conference,letterpaper]{IEEEtran}

\usepackage[table]{xcolor}
\usepackage[utf8]{inputenc}
\usepackage[T1]{fontenc}
\usepackage{url}
\usepackage{ifthen}

\usepackage{subcaption}
\usepackage[cmex10]{amsmath} % Use the [cmex10] option to ensure complicance
\usepackage{graphicx, amssymb, tikz, multicol, amsthm} % Required for inserting images
\usetikzlibrary{patterns}
\usepackage{mathdots}
\usepackage[colorlinks=true,linkcolor=black,anchorcolor=black,citecolor=black,filecolor=black,menucolor=black,runcolor=black,urlcolor=black]{hyperref}
\usepackage{booktabs}
\usepackage{float}
\usepackage{multicol}
\usepackage{caption}
\usepackage{gnuplottex}
\usepackage{cite}

\newtheorem{thm}{Theorem}
\newtheorem{cor}{Corollary}
\newtheorem{lem}{Lemma}

\theoremstyle{definition}

\newtheorem{defn}{Definition}

\newcommand{\cA}{\mathcal{A}}
\newcommand{\cB}{\mathcal{B}}
\newcommand{\cR}{\mathcal{R}}
\newcommand{\cS}{\mathcal{S}}
\newcommand{\supp}{\mathrm{supp}}

\newcommand{\Gsmall}{\mathsf{GASP}_{\mathrm{small}}}
\newcommand{\Gbig}{\mathsf{GASP}_{\mathrm{big}}}
\newcommand{\Gr}{\mathsf{GASP}_{r}}
\newcommand{\GASP}{\mathsf{GASP}}
\newcommand{\Grasp}{\mathsf{GRASP}}
\newcommand{\DOG}{\mathsf{DOG}}
\newcommand{\CAT}{\mathsf{CAT}}
\newcommand{\SHIFT}{\mathsf{SHIFT}}
\newcommand{\COVER}{\mathsf{COVER}}
\newcommand{\PG}{\mathsf{PoleGap}}

\newcommand{\Npre}{N^{\mathrm{pre}}}
\newcommand{\Nint}{N_{\mathrm{int}}^*}
\DeclareMathOperator{\gap}{gap}

\definecolor{DarkGreen}{rgb}{0.1,0.5,0.1}
\definecolor{DarkRed}{rgb}{0.5,0.1,0.1}
\definecolor{DarkBlue}{rgb}{0.1,0.1,0.5}
\definecolor{DarkPurple}{rgb}{0.5,0.2,0.5}
\definecolor{DarkTurquoise}{rgb}{0.1,0.5,0.5}
\definecolor{DarkGray}{RGB}{75,75,75}

\definecolor{Orange}{RGB}{255, 132, 0}
\definecolor{Yellow}{RGB}{245, 220, 50}
\definecolor{Green}{RGB}{90,190,0}
\definecolor{Blue}{RGB}{0,150,250}
\definecolor{Indigo}{RGB}{40,0,250}
\definecolor{Violet}{RGB}{160,0,250}
\definecolor{Pink}{RGB}{240,0,250}
\definecolor{pinegreen}{rgb}{0.0, 0.47, 0.44}

\begin{document}
\title{Secure Distributed Matrix Multiplication \\ with Precomputation}

\author{%
  \IEEEauthorblockN{Ryann Cartor\IEEEauthorrefmark{1},
                    Rafael G. L.~D'Oliveira\IEEEauthorrefmark{1},
                    Salim El Rouayheb\IEEEauthorrefmark{2},
                    Daniel Heinlein \IEEEauthorrefmark{3}, \\
                    David Karpuk \IEEEauthorrefmark{4},
                    Freeman Slaughter \IEEEauthorrefmark{5}, and
                    Alex Sprintson \IEEEauthorrefmark{6}}
  \IEEEauthorblockA{\IEEEauthorrefmark{1}%
                   School of Mathematical and Statistical Sciences, Clemson University, Clemson, SC, USA,
                    \{rcartor,rdolive\}@clemson.edu}
  \IEEEauthorblockA{\IEEEauthorrefmark{2}%%%%Salim
                   ECE Department, Rutgers University,
                    New Brunswick, NJ, USA,
                    salim.elrouayheb@rutgers.edu}
  \IEEEauthorblockA{\IEEEauthorrefmark{3}%%%%% Daniel
                    Independent Researcher, San Francisco, USA, contact@danielheinlein.com}
    \IEEEauthorblockA{\IEEEauthorrefmark{4}%%%%%%% Dave
                    QMill, Espoo, Finland, david.karpuk@qmill.com}
    \IEEEauthorblockA{\IEEEauthorrefmark{5}%%%%% Freeman
                    Department of Mathematics \& Statistics,  University of South Florida, Tampa, FL, USA,
                   fslaughter@usf.edu}
    \IEEEauthorblockA{\IEEEauthorrefmark{6}%%%%% Alex, Texas A&M
                    Department of Electrical and Computer Engineering, George Mason University, Fairfax, VA, USA
                   asprints@gmu.edu}
}

\maketitle

%%%%%%

\begin{abstract}

We consider secure distributed matrix multiplication, in which a user wishes to compute the product of two matrices with the assistance of honest-but-curious servers, any $T$ of which may collude, using the outer-product partition. 
The matrices are split into $K$ and $L$ blocks which are then hidden by random masks, and the user is assumed to have precomputed the products of the random matrices in advance. 
We then develop methods to leverage this precomputation to reduce the number of servers required.

Polynomial codes for this problem are designed through degree tables, whose distinct entries count the servers. 
We show that precomputation can reduce the minimum number of servers required by designing degree tables that are specifically optimized for the precomputation setting.
In particular, for Gap Additive Secure Polynomial ($\Gr$) codes, we give a precise formula, and show that it is monotone in the chain length $r$ -- so that only the extremes $\Gsmall$ and $\Gbig$ matter -- and prove lower bounds that $\Gsmall$ attains in several regimes. 
Two consequences are that any fraction of colluding servers below one can be tolerated, compared to the threshold of one half without precomputation, and that the order of the total time complexity decreases when the colluders are a fixed fraction of the servers. 
We also show that no existing construction utilizes fewer servers than the better of $\Gsmall$ and $\Gbig$ once the mask products are precomputed.

We then introduce Gap-Reduced Additive Secure Polynomial ($\Grasp$) codes, which retain a single interval for the mask degrees, while carefully selecting degrees that represent the information supports. 
For $K = T = 4$ and every $L \geq 3$, we provide an explicit $\Grasp$ code that uses one server fewer than every $\Gr$ code, and is within two servers of optimal. 
For large $T$, the configuration becomes an integer packing problem that is independent of $T$, and the number of servers that $\Grasp$ codes outperform $\Gr$ by grows quadratically in $K$ and $L$.
\end{abstract}

\section{Introduction}

We consider the problem of secure distributed matrix multiplication (SDMM) where a user has two matrices, $A$ and $B$, and wishes to compute their product, $A \cdot B$, with the help of $N$ servers, without leaking any information about either $A$ or $B$ to any server. We consider the case where the servers are honest but curious, and assume that at most $T$ of them collude to try to deduce information about $A$ or $B$.

Various performance metrics have been studied in the literature for this problem, including download cost~\cite{SDMM1/TandonChang2018}, total communication cost~\cite{kakar2020uplinkdownlink,upload_vs_download,GASP/DOliveiraRK20}, and total time complexity~\cite{rafael_note}. 
For polynomial codes with fixed matrix partitioning, these performance metrics are equivalent to minimizing the minimum number of servers needed, $N$, often called the scheme's recovery threshold \cite{field_trace}. 
In this paper, we consider the outer product partitioning given by
\begin{equation}\label{AB_partition}
A=\begin{bmatrix}
    A_1\\\vdots\\ A_K
\end{bmatrix},\quad B=\begin{bmatrix}
    B_1&\cdots&B_L
\end{bmatrix}
\end{equation}
so that
\begin{equation*}
AB=\begin{bmatrix}
    A_1B_1&\cdots &A_1B_L\\
    \vdots&\ddots&\vdots\\
    A_KB_1&\cdots&A_KB_L
\end{bmatrix}
\end{equation*}
where computing $AB$ is equivalent to each submatrix $A_kB_\ell$, $k\in [K], l\in [L]$.

In this paper we focus on computation using \emph{polynomial codes}.
That is, to compute each submatrix $A_kB_\ell$, we create a polynomial $h(x) = f(x) \cdot g(x)$, where the coefficients represent the products $A_kB_\ell$. Each server then receives evaluations of $f$ and $g$, and multiplies them to obtain an evaluation of $h$. The polynomials $f$ and $g$ are designed so that any group of $T$ colluding servers cannot obtain information about $A$ or $B$, while the user can reconstruct $AB$ from the evaluations of $h$.

The polynomial construction can be performed by designing an appropriate \emph{degree table}~\cite{GASP/DOliveiraRK20, DegTable/DOliveiraRHK21}. In particular,  \cite[Theorem~1]{GASP/DOliveiraRK20} establishes that if the degree table meets specific criteria, then the polynomial code is both decodable and secure against $T$ colluding servers, with $N$ corresponding to the number of unique terms in the table. The predominant challenge is then to construct degree tables that minimize $N$. The most efficient known schemes for the outer product partitioning are a family of schemes, Gap Additive Secure Polynomial ($\Gr$) \cite{GASP/DOliveiraRK20}, parameterized by a parameter $r$, which is referred to as the \emph{chain length}.
\footnote{The detailed definition of $\Gr$ appears in Section~\ref{sec:gasp-precomputation} below.}

This paper focuses on settings where the user can precompute products of the random matrices and develop methods to leverage this precomputation to reduce the number of servers required. The paper is motivated by
several practical settings where it is possible to perform a certain amount of computation in advance, e.g., during server downtime. 
The Beaver triple~\cite{Beaver} is a relevant example of how precomputation can be beneficial in secure matrix multiplications.
The advantage of using precomputation has been observed by Mital et~al.\ in the context of inner product computation~\cite{inner_product}. They note that the product of the random matrices can be computed in advance and stored at the user to facilitate the secure matrix multiplication process upon request.
Precomputation has also been used in the context of Secure Multiparty Computation (MPC) (see, e.g.,~\cite{10.1007/978-3-642-32009-5_38,10.1007/978-3-319-39555-5_18,10.1007/978-3-642-32009-5_40,10.1007/978-3-030-36030-6_14}) and references therein. Damgård et~al.\ proposed an MPC scheme that includes a precomputation phase (referred to as offline or preprocessing phase) to generate correlated randomness between the parties. This phase is independent of both the inputs and the function computed by the MPC algorithm. Such protocols have been shown to have significant advantages in practical applications~\cite{10.1007/978-3-642-32009-5_40} due to offloading of heavy computation to the offline phase.

\begin{figure*}[!t]
 \includegraphics[trim={2cm 4.9in 1cm 4.2in}, clip, width=\textwidth]{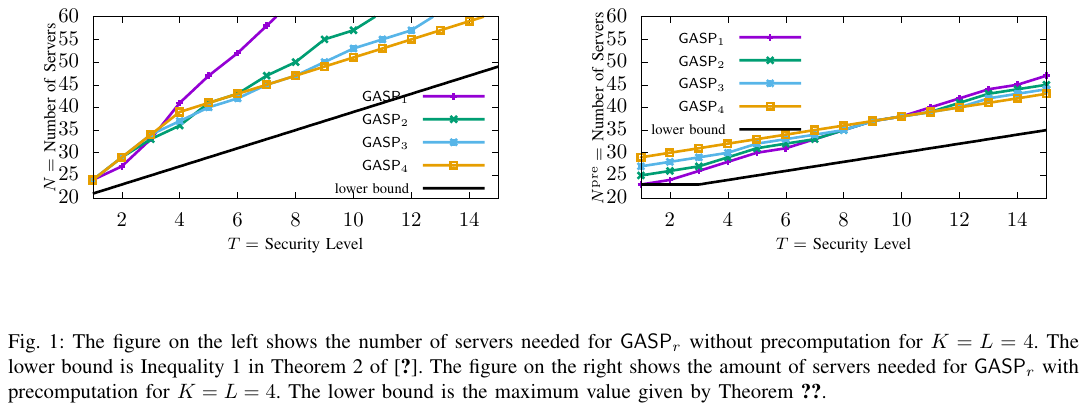} 
  \vspace*{-3mm}
 \caption{The figure on the left shows the number of servers needed for $\Gr$ without precomputation for $K=L=4$. The lower bound is Inequality~1 in Theorem~2 of~\cite{DegTable/DOliveiraRHK21}. The figure on the right shows the amount of servers needed for $\Gr$ with precomputation for $K=L=4$. The lower bound is the maximum value given by Theorem~\ref{thm:PrecompBound}. }
  \label{fig:KL_fixed:T_grows}
\end{figure*}

This paper makes the following contributions. First,
we show that precomputations by the user can reduce the minimum number of servers required. The key insight is that degrees in the lower right corner of the degree table correspond to coefficients of the form $R_t S_{t'}$, which are independent of the matrices $A$ and $B$, and can be precomputed by the user. The optimization problem shifts to designing degree tables that fulfill the original conditions while minimizing the number of distinct terms, excluding those in the lower right corner.

Precomputation does more than reduce the number of servers required by an existing scheme. It changes which degrees contribute to the server count, and therefore changes the degree tables that are desirable. A construction designed to minimize the number of servers without precomputation may concentrate repeated degrees in the lower-right block of the degree table. When this block is precomputed, those repetitions no longer reduce the number of servers needed for the online computation. Thus, rather than first choosing a scheme for the setting without precomputation and then precomputing its random–random products, we can design degree tables that are specifically optimized for the precomputation setting.

\subsection{Main Contributions}

\begin{itemize}

    \item We establish the number of servers needed for $\Gr$ in the precomputation setting in Theorem~\ref{thm:GasprServers1}.

    \item We present lower bounds for the degree table construction with precomputation in Theorem~\ref{thm:PrecompBound}, and use them to prove the optimality of some of our schemes.

    \item We show that precomputation allows us to tolerate any fixed percentage (below $100 \%$) of collusions relative to the total number of servers. Without precomputation, the percentage of colluding servers cannot exceed 50\%.

    \item We show, in Theorems~\ref{thm:complexity_no_precomp} and~\ref{thm:complexity_with_precomp}, that precomputation reduces the order of the time complexity for the case where the number of colluding servers is a fixed percentage of the total number of servers.

    \item We introduce Gap-Reduced Additive Secure Polynomial ($\Grasp$) codes, which allow information supports beyond those used in $\Gr$. We present common-interval, split, and unified constructions, with the unified construction including $\Gr$ and the split construction as special cases.

    \item We show in Theorem~\ref{thm:span_vs_gasp} that, for sufficiently large $T$, $\Grasp$ codes require fewer servers than every $\Gr$ code when $m=\min\{K,L\}\geq 4$, or when $m=3$ and $M=\max\{K,L\}$ is even. The improvement can grow quadratically in the partition parameters.

\end{itemize}

\subsection{Related Work}

Our method for constructing coded systems for Secure Distributed Matrix Multiplication (SDMM) follows the general strategy of \emph{polynomial codes}, introduced first in~\cite{PC/YuMA2017} for distributed matrix multiplication with stragglers and later adapted by~\cite{SDMM1/TandonChang2018} for secure systems. This construction was later improved upon in~\cite{SDMM2/KakarES2019, GASP/DOliveiraRK20, DegTable/DOliveiraRHK21}, the latter two of which introduced the $\GASP$ codes that vastly improved upon previous constructions in terms of rate. These codes were recently generalized to certain algebraic curves in~\cite{AGSDMM/MakkonenSH2023}, which were shown to offer modest improvements over the $\GASP$ codes depending on the parameter ranges.

If we consider other partitions of $A$ and $B$, the Discrete Fourier Transform codes of~\cite{inner_product} are the state-of-the-art for the inner product partition, in which $A$ is partitioned vertically and $B$ horizontally. The inner product partition was also studied in~\cite{hermitian} in which the authors use codes constructed over algebraic curves to decrease the required size of the base field. In~\cite{field_trace} the authors used the field trace to decrease total communication cost without decreasing the number of work nodes. For the \emph{grid partition} in which both matrices are partitioned in both directions, many schemes are known~\cite{oliver, root_of_unity, karpuk2024modular} but none is clearly superior in all parameter regimes.

The aforementioned work all focuses on minimizing the number of worker nodes employed by the scheme, but several other works consider other objective functions and study trade-offs between upload cost, download cost, encoding and decoding complexity, and latency~\cite{bivariate, rawad_adaptive, rawad_latency, upload_vs_download, systematic_private}, to name a few. SDMM also bears similarities to Private Information Retrieval, a connection that was exploited in~\cite{jafar_batch, jafar_batch2, jafar_cross, jafar_xsecure_tprivate} to give schemes for batch matrix multiplication in many setups of interest. In the other direction, techniques from SDMM have also found applications in private information retrieval~\cite{kazemi2022degree}.

\subsection{Some Notation and Basic Definitions}

\begin{defn}
The \emph{support} of a polynomial $h(x) = \sum_i a_ix^i$ is $\supp(h) = \{i\ |\ a_i\neq 0\}$.
\end{defn}

Using generalized Vandermonde matrices, by appropriately choosing the evaluation points,
one can interpolate $h(x)$ and therefore recover all of its coefficients with $|\supp(h)|$ evaluations over a sufficiently large field $\mathbb{F}$.

\begin{defn}
For real numbers $x$ and $y$, we denote by $[x:y]$ the set of all integers in the interval $[x,y]$. If $y < x$ then we define $[x:y] = \emptyset$. We define $[n]=[0:n]=\{0,1,\ldots,n\}$ for a nonnegative integer $n$, and $[n] = \emptyset$ if $n < 0$. Lastly, we define $a+B+c \cdot D=\{a+b+c \cdot d : b \in B, d \in D\}$ for $a,c \in \mathbb{Z}$, $B \subseteq \mathbb{Z}$, and $D \subseteq \mathbb{Z}$.
\end{defn}

\subsection{Paper Outline}

 The rest of the paper is structured as follows. First, in Section~\ref{sec:outline}, we present background on polynomial codes for secure matrix multiplication and the concept of \emph{degree tables}, and we define polynomial codes with precomputation. In Section~\ref{sec:gasp-precomputation}, we analyze the performance of $\Gr$ codes with precomputation. In Section~\ref{sec:bounds}, we establish bounds on the number of servers and the degree of tolerance to collusion in the presence of precomputation, and we also show that precomputation reduces time complexity. In Section~\ref{sec:GRASP}, we introduce $\Grasp$ codes, the main contribution of this paper.
Next, in Section~\ref{sec:comparison}, we compare the performance of our scheme with other schemes proposed in the literature. 
Finally, we present our conclusions in Section~\ref{sec:conclusion}. The proofs of all presented lemmas and theorems appear in the Appendix.

\section{Polynomial Codes for Secure Distributed Matrix Multiplication}\label{sec:outline}

\subsection{Polynomial Codes without Precomputation}

Let $A$ and $B$ be the two matrices, over some finite field $\mathbb{F}_q$, which we want to multiply. We partition the matrices as in~\eqref{AB_partition}, and construct a polynomial code to distribute their multiplication as follows. We choose two polynomials $f(x)$ and $g(x)$ that encode the blocks of $A$ and $B$, respectively. The coefficients of their product $h(x) = f (x) \cdot g(x)$ encode the submatrices $A_kB_\ell$. Thus, interpolating $h(x)$ suffices to reconstruct the product $AB$. We then utilize $N$ servers to compute the evaluations $h(a_1),\ldots, h(a_N)$ for certain $a_1,\ldots, a_N \in \mathbb{F}$.

The polynomials $f(x)$ and $g(x)$ are constructed so that any $T$-subset of evaluations reveals no information about $A$ or $B$ (referred to as $T$-security), but so that the user can reconstruct all of $AB$ given all $N$ evaluations (decodability). Precise information-theoretic descriptions of these properties can be found in, e.g.,~\cite[Definition~2]{GASP/DOliveiraRK20}.

The polynomials $f(x)$ and $g(x)$ are sums of ``information'' and ``random'' polynomials, defined as follows:
\begin{align}\label{eq:f_g_polys}
\begin{split}
f_I(x) &= \sum_{k = 1}^K A_kx^{\alpha_k},\quad f_R(x) = \sum_{t = 1}^TR_tx^{\alpha_{K+t}} \\
f(x) &= f_I(x) + f_R(x) \\
g_I(x) &= \sum_{\ell = 1}^L B_\ell x^{\beta_\ell},\quad g_R = \sum_{t = 1}^T S_t x^{\beta_{L + t}} \\
g(x) &= g_I(x) + g_R(x)
\end{split}
\end{align}
where $R_t$ and $S_t$ are random matrices chosen for privacy, and $\alpha_1,\ldots, \alpha_{K+T}, \beta_1,\ldots,\beta_{L+T}$ determine the scheme.  We abbreviate $\alpha_I = (\alpha_1,\ldots,\alpha_K)$ and $\alpha_R = (\alpha_{K +1},\ldots,\alpha_{K+T})$, and write $\alpha = (\alpha_I|\alpha_R) = (\alpha_1,\ldots,\alpha_{K+T})$
for their concatenation.  We similarly define $\beta_I$, $\beta_R$, and $\beta$.  It is straightforward to see that $\alpha_I = \supp(f_I)$, $\alpha_R = \supp(f_R)$, and $\alpha = \supp(f)$, and similarly $\beta_I = \supp(g_I)$, etc. We often treat $\alpha$ and $\beta$ as vectors and sets simultaneously.

With $f(x)$, $g(x)$, and $h(x)$ as above, the scheme uses $N = |\supp(h)|$ servers. This set can be expressed as
\begin{align*}
\begin{split}
    \supp(h) =& \alpha + \beta \\
    =& (\alpha_I + \beta_I) \cup (\alpha_I + \beta_R) \cup (\alpha_R + \beta_I) \cup (\alpha_R + \beta_R).
\end{split}
\end{align*}
With the goal of minimizing the number of servers needed, we are tasked with choosing $\alpha$ and $\beta$ to minimize $N$.

\subsection{Degree Tables}

The degree table~\cite{GASP/DOliveiraRK20,DegTable/DOliveiraRHK21} of $h(x)$, depicted in Table~\ref{tab:DegTable}, shows the exponents in $h(x)$ as a function of $\alpha$ and $\beta$. In~\cite[Theorem~1]{GASP/DOliveiraRK20}, it is shown that if the degree table satisfies the following conditions: (i)~the numbers in the red block are unique in the table and (ii)~numbers in the green/blue block are pairwise distinct, respectively, then there exist evaluation points such that the polynomial code in Equation~\eqref{eq:f_g_polys} is decodable and $T$-secure. Furthermore, the minimum number of servers, $N$, is the number of distinct terms in the table. Thus, the main question in the literature has been how to construct a degree table satisfying conditions (i) and (ii) so as to minimize the number of servers, $N$. 
For this task, the $\Gr$ family of codes is currently the best-known scheme in the literature, where  $r$ is the number of consecutive degrees placed in each block of length $K$.
%are given by a family of schemes known as $\Gr$. 
The degree tables for $\Gr$ codes are illustrated in Figures~\ref{fig:KL_fixed:T_grows} (a) and (b).

\subsection{Polynomial Codes with Precomputation}

\begin{table}
\setlength{\tabcolsep}{3pt}
\centering
\resizebox{.48\textwidth}{!}{
\begin{tabular}{c|ccc|ccc}

 & $\beta_1$ & $\!\!\cdots\!\!$ & $\beta_L$ &  \cellcolor{blue!25} $\beta_{L\!+\!1}$ & \cellcolor{blue!25} $\!\!\cdots\!\!$ & \cellcolor{blue!25} $\beta_{L\!+\!T}$  \\
\toprule
$\alpha_1$ & \cellcolor{red!25} $\alpha_1 \!+\! \beta_1$ & \cellcolor{red!25} $\!\!\cdots\!\!$ & \cellcolor{red!25} $\alpha_1 \!+\! \beta_L$ & $\alpha_1 \!+\! \beta_{L\!+\!1}$ & $\!\!\cdots\!\!$ & $\alpha_1 \!+\! \beta_{L\!+\!T}$  \\
$\vdots$ & \cellcolor{red!25} $\vdots$ & \cellcolor{red!25} $\!\!\ddots\!\!$ & \cellcolor{red!25} $\vdots$ & $\vdots$ & $\!\!\ddots\!\!$ & $\vdots$  \\
$\alpha_K$ & \cellcolor{red!25} $\alpha_K \!+\! \beta_1$ & \cellcolor{red!25} $\!\!\cdots\!\!$ & \cellcolor{red!25} $\alpha_K \!+\! \beta_L$ & $\alpha_K \!+\! \beta_{L\!+\!1}$ & $\!\!\cdots\!\!$ & $\alpha_K \!+\! \beta_{L\!+\!T}$  \\
\midrule
\cellcolor{green!25} $\alpha_{K\!+\!1}$ &  $\alpha_{K\!+\!1} \!+\! \beta_1$ & $\!\!\cdots\!\!$ &  $\alpha_{K\!+\!1} \!+\! \beta_L$ & $\alpha_{K\!+\!1} \!+\! \beta_{L\!+\!1}$ & $\!\!\cdots\!\!$ & $\alpha_{K\!+\!1} \!+\! \beta_{L\!+\!T}$ \\
\cellcolor{green!25} $\vdots$ &  $\vdots$ & $\!\!\ddots\!\!$ &  $\vdots$ & $\vdots$ & $\!\!\ddots\!\!$ & $\vdots$ \\
\cellcolor{green!25} $\alpha_{K\!+\!T}$ &  $\alpha_{K\!+\!T} \!+\! \beta_1$ & $\!\!\cdots\!\!$ &  $\alpha_{K\!+\!T} \!+\! \beta_L$ & $\alpha_{K\!+\!T} \!+\! \beta_{L\!+\!1}$ & $\!\!\cdots\!\!$ & $\alpha_{K\!+\!T} \!+\! \beta_{L\!+\!T}$ \\
\bottomrule
\end{tabular}
    }
\caption{The Degree Table. The $\alpha_i$'s and $\beta_i$'s are the exponents of the polynomials $f(x)$ and $g(x)$ in~\eqref{eq:f_g_polys} used to  encode  $A$ and $B$, respectively. The table entries are the  monomial  degrees in  $f(x)\cdot g(x)$. The problem is to choose the degrees $\alpha_i$'s and $\beta_i$'s to minimize the number of distinct entries in the table subject to decodability %, the red block entries must be unique in the table;
and  $T$-security. %, all green/blue block entries must be pairwise distinct.
% In Theorem~1 of~\cite{GASP/DOliveiraRK20} it is shown any base field $\mathbb{F}_q$ can be extended to guarentee (i) and (ii). Here we assume that $\mathbb{F}_q$ already has the required size.
}\label{tab:DegTable}
\end{table}

The unique entries in the lower right corner of the degree table in Table~\ref{tab:DegTable} correspond precisely to $\supp(f_Rg_R) = \alpha_R + \beta_R$, which is independent of the matrices $A$ and $B$. Consequently, the user can precompute the products $R_tS_{t'}$ in advance. Once $A$ and $B$ are available, they only need to outsource the remaining computation. This approach effectively eliminates the need to consider the term $\alpha_R + \beta_R$ when computing $|\supp(h)|$, reducing the server requirement to $\Npre = |(\alpha_I+\beta_I)\cup(\alpha_I+\beta_R)\cup(\alpha_R+\beta_I)|$. In this work, we consider the case of studying codes specifically for SDMM with precomputation, where the objective of minimizing $N$ has been replaced by that of minimizing $\Npre$.

\begin{figure*}[!t]
\setlength{\tabcolsep}{0.33em}
% \renewcommand{\arraystretch}{0.9}
% \subfloat[$r=1$, i.e. $\mathsf{GASP}_{\text{small}}$, $N=41$\label{tab:KLT4_example_r1}]{
\begin{subtable}{0.24\textwidth}
\resizebox{\columnwidth}{!}{\begin{tabular}{c|cccc|cccc}
  & 0& 4& 8&12&\cellcolor{blue!25}16&\cellcolor{blue!25}17&\cellcolor{blue!25}18&\cellcolor{blue!25}19\\
\toprule
 0& \cellcolor{red!25}0& \cellcolor{red!25}4& \cellcolor{red!25}8&\cellcolor{red!25}12&\cellcolor{gray!25}16&\cellcolor{gray!25}17&\cellcolor{gray!25}18&\cellcolor{gray!25}19\\
 1& \cellcolor{red!25}1& \cellcolor{red!25}5& \cellcolor{red!25}9&\cellcolor{red!25}13&17&18&19&\cellcolor{gray!25}20\\
 2& \cellcolor{red!25}2& \cellcolor{red!25}6&\cellcolor{red!25}10&\cellcolor{red!25}14&18&19&20&\cellcolor{gray!25}21\\
 3& \cellcolor{red!25}3& \cellcolor{red!25}7&\cellcolor{red!25}11&\cellcolor{red!25}15&19&20&21&\cellcolor{gray!25}22\\
\midrule
16&16&20&\cellcolor{gray!25}24&\cellcolor{gray!25}28&\cellcolor{gray!25}32&\cellcolor{gray!25}33&\cellcolor{gray!25}34&\cellcolor{gray!25}35\\
\cellcolor{green!25}20&20&24&28&32&\cellcolor{gray!25}36&\cellcolor{gray!25}37&\cellcolor{gray!25}38&\cellcolor{gray!25}39\\
\cellcolor{green!25}24&24&28&32&36&\cellcolor{gray!25}40&\cellcolor{gray!25}41&\cellcolor{gray!25}42&4\cellcolor{gray!25}3\\
\cellcolor{green!25}28&28&32&36&40&\cellcolor{gray!25}44&\cellcolor{gray!25}45&\cellcolor{gray!25}46&\cellcolor{gray!25}47\\
\bottomrule
\end{tabular}}
\caption{$r\!\!=\!1$, $N_1\!\!=\!41$}\label{tab:KLT4_example_r1}
\end{subtable}
% }
\begin{subtable}{0.24\textwidth}
% \subfloat[$r=2$, i.e. $\mathsf{GASP}_{\text{medium}}$, $N=36$\label{tab:KLT4_example_r2}]{
\resizebox{\columnwidth}{!}{\begin{tabular}{c|cccc|cccc}
  & 0& 4& 8&12&\cellcolor{blue!25}16&\cellcolor{blue!25}17&\cellcolor{blue!25}18&\cellcolor{blue!25}19\\
\toprule
 0& \cellcolor{red!25}0& \cellcolor{red!25}4& \cellcolor{red!25}8&\cellcolor{red!25}12&\cellcolor{gray!25}16&\cellcolor{gray!25}17&1\cellcolor{gray!25}8&\cellcolor{gray!25}19\\
 1& \cellcolor{red!25}1& \cellcolor{red!25}5& \cellcolor{red!25}9&\cellcolor{red!25}13&17&18&19&\cellcolor{gray!25}20\\
 2& \cellcolor{red!25}2& \cellcolor{red!25}6&\cellcolor{red!25}10&\cellcolor{red!25}14&18&19&20&\cellcolor{gray!25}21\\
 3& \cellcolor{red!25}3& \cellcolor{red!25}7&\cellcolor{red!25}11&\cellcolor{red!25}15&19&20&21&\cellcolor{gray!25}22\\
\midrule
\cellcolor{green!25}16& 16& 20&\cellcolor{gray!25}24&\cellcolor{gray!25}28&\cellcolor{gray!25}32&\cellcolor{gray!25}33&\cellcolor{gray!25}34&\cellcolor{gray!25}35\\
\cellcolor{green!25}17& 17& 21&\cellcolor{gray!25}25&\cellcolor{gray!25}29& 33& 34& 35&\cellcolor{gray!25}36\\
\cellcolor{green!25}20& 20& 24& 28& 32& 36&\cellcolor{gray!25}37&\cellcolor{gray!25}38&\cellcolor{gray!25}39\\
\cellcolor{green!25}21& 21& 25& 29& 33& 37& 38& 39&\cellcolor{gray!25}40\\
\bottomrule
\end{tabular}}
\caption{$r\!\!=\!2$, $N_2\!\!=\!36$}\label{tab:KLT4_example_r2}
% }
\end{subtable}
% \subfloat[$r=3$, $N=37$\label{tab:KLT4_example_r3}]{
\begin{subtable}{0.24\textwidth}
\resizebox{\columnwidth}{!}{\begin{tabular}{c|cccc|cccc}
  & 0& 4& 8&12&\cellcolor{blue!25}16&\cellcolor{blue!25}17&\cellcolor{blue!25}18&\cellcolor{blue!25}19\\
\toprule
 0& \cellcolor{red!25}0& \cellcolor{red!25}4& \cellcolor{red!25}8&\cellcolor{red!25}12&\cellcolor{gray!25}16&\cellcolor{gray!25}17&\cellcolor{gray!25}18&\cellcolor{gray!25}19\\
 1& \cellcolor{red!25}1& \cellcolor{red!25}5& \cellcolor{red!25}9&\cellcolor{red!25}13&17&18&19&\cellcolor{gray!25}20\\
 2& \cellcolor{red!25}2& \cellcolor{red!25}6&\cellcolor{red!25}10&\cellcolor{red!25}14&18&19&20&\cellcolor{gray!25}21\\
 3& \cellcolor{red!25}3& \cellcolor{red!25}7&\cellcolor{red!25}11&\cellcolor{red!25}15&19&20&21&\cellcolor{gray!25}22\\
\midrule
16&16&20&\cellcolor{gray!25}24&\cellcolor{gray!25}28&\\%\cellcolor{black}32&\cellcolor{gray!25}33&\cellcolor{gray!25}34&\cellcolor{gray!25}35\\
\cellcolor{green!25}20&20&24&28&\cellcolor{gray!25}32&\\%\cellcolor{gray!25}%36&\cellcolor{gray!25}37&\cellcolor{gray!25}38&\cellcolor{gray!25}39\\
\cellcolor{green!25}24&24&28&32&\cellcolor{gray!25}36&\\%\cellcolor{gray!25}40&\cellcolor{gray!25}41&\cellcolor{gray!25}42&4\cellcolor{gray!25}3\\
\cellcolor{green!25}28&28&32&36&\cellcolor{gray!25}40&\\%\cellcolor{gray!25}44&\cellcolor{gray!25}45&\cellcolor{gray!25}46&\cellcolor{gray!25}47\\
\bottomrule
\end{tabular}}\caption{$r\!\!=\!1$, $\Npre_1\!\!=\!28$}\label{tab:KLT4_example_r1-pre}
\end{subtable}
% }
% \subfloat[$r=4$, i.e. $\mathsf{GASP}_{\text{big}}$, $N=39$\label{tab:KLT4_example_r4}]{
\begin{subtable}{0.24\textwidth}
\resizebox{\columnwidth}{!}{\begin{tabular}{c|cccc|cccc}
  & 0& 4& 8&12&\cellcolor{blue!25}16&\cellcolor{blue!25}17&\cellcolor{blue!25}18&\cellcolor{blue!25}19\\
\toprule
 0& \cellcolor{red!25}0& \cellcolor{red!25}4& \cellcolor{red!25}8&\cellcolor{red!25}12&\cellcolor{gray!25}16&\cellcolor{gray!25}17&1\cellcolor{gray!25}8&\cellcolor{gray!25}19\\
 1& \cellcolor{red!25}1& \cellcolor{red!25}5& \cellcolor{red!25}9&\cellcolor{red!25}13&17&18&19&\cellcolor{gray!25}20\\
 2& \cellcolor{red!25}2& \cellcolor{red!25}6&\cellcolor{red!25}10&\cellcolor{red!25}14&18&19&20&\cellcolor{gray!25}21\\
 3& \cellcolor{red!25}3& \cellcolor{red!25}7&\cellcolor{red!25}11&\cellcolor{red!25}15&19&20&21&\cellcolor{gray!25}22\\
\midrule
\cellcolor{green!25}16& 16& 20&\cellcolor{gray!25}24&\cellcolor{gray!25}28&\\%\cellcolor{gray!25}32&\cellcolor{gray!25}33&\cellcolor{gray!25}34&\cellcolor{gray!25}35\\
\cellcolor{green!25}17& 17& 21&\cellcolor{gray!25}25&\cellcolor{gray!25}29&\\% 33& 34& 35&\cellcolor{gray!25}36\\
\cellcolor{green!25}20& 20& 24& 28& \cellcolor{gray!25}32& \\%36&\cellcolor{gray!25}37&\cellcolor{gray!25}38&\cellcolor{gray!25}39\\
\cellcolor{green!25}21& 21& 25& 29&\cellcolor{gray!25} 33& \\%37& 38& 39&\cellcolor{gray!25}40\\
\bottomrule
\end{tabular}}
\caption{$r\!\!=\!2$,  $\Npre_2\!\!=\!29$}\label{tab:KLT4_example_r2-pre}
\end{subtable}
% }
\caption{The degree tables for $\GASP_2$, and $\GASP_1$ when $K=L=T=4$.  The gray regions represent the first time a number appears in a degree table, and determine 
%denotes 
the number of distinct nodes needed. It is shown in~\cite{DegTable/DOliveiraRHK21} that $\GASP_2$ is the best choice for these parameters without precomputation (Tables~\ref{tab:KLT4_example_r1},~\ref{tab:KLT4_example_r2}). When we consider precomputation, $\GASP_1$ is better (Tables~\ref{tab:KLT4_example_r1-pre},~\ref{tab:KLT4_example_r2-pre}).}
\label{tab:KLT4_example}
\end{figure*}

As a motivating example, we note that for parameters $K=L=T=4$, $\GASP_2$ is the superior scheme without precomputation, utilizing $N_2 = 36$ servers as depicted in the second table of Figure~\ref{tab:KLT4_example}. This contrasts with $\GASP_1$, which requires $N_1 = 41$ servers, detailed in the first table of the same figure. However, introducing precomputation reverses their performance rankings. Specifically, $\GASP_1$ becomes more efficient, requiring only $\Npre_1 = 28$ servers, compared to $\GASP_2$, which needs $\Npre_2 = 29$. These comparisons are illustrated in the right two tables of Figure~\ref{tab:KLT4_example}.

However, the efficiency of $\GASP_1$ in precomputation scenarios is not universal. For example, with parameters $K=L=4$ and $T=11$, $\GASP_1$ requires $\Npre_1 = 40$ servers, while $\GASP_4$ proves to be more efficient, requiring only $\Npre_4 = 39$ servers.
We note that both $\GASP_1$ and $\Gr$ with $r = \min\{K,T\}$  were first introduced in~\cite{GASP/DOliveiraRK20} under the names $\Gsmall$ and $\Gbig$.

\section{\texorpdfstring{$\Gr$}{GASPr} with Precomputation}\label{sec:gasp-precomputation}

We recall the definition of the polynomial code $\Gr$ from~\cite{DegTable/DOliveiraRHK21}.  As these codes represent the state-of-the-art for a large swath of parameter ranges in the case of no precomputation, it is natural to analyze their performance in the precomputation setting.

Given two matrices $A$ and $B$ with the outer product partition, to define $f(x)$ and $g(x)$ as in~\eqref{eq:f_g_polys} it suffices to define $\alpha_I$, $\alpha_R$, $\beta_I$, and $\beta_R$.  We first set
\begin{align}
    \begin{split}
        \alpha_I &= (0,1,\ldots,K-1) = [K-1] \\
        \beta_I &= (0,K,\ldots,(L-1)K) = K\cdot [L-1]
    \end{split}
\end{align}
To define $\alpha_R$ and $\beta_R$ we first set
\[
\beta_R = (KL, KL + 1,\ldots, KL + T-1) = KL + [T-1].
\]
We now choose a positive integer $r\leq \min\{K,T\}$ and use Euclidean division to write $T = Ur + r_0$ for $U = \left\lfloor T/r\right\rfloor$ and $0\leq r_0 < r$.  We then set the first $T$ integers in the sequence
\begin{align*}
    \alpha_R &= KL + \bigcup_{u = 1}^{\infty}\left((u-1)K + [r-1]\right) \\
    &= KL + \bigcup_{u = 1}^U\left((u-1)K + [r-1]\right) \cup \left(UK + [r_0-1]\right)
\end{align*}
When $r = 1$ we recover the polynomial code $\Gsmall$ and when $r =\min\{K,T\}$ we recover $\Gbig$, both of which first appeared in~\cite{GASP/DOliveiraRK20}.

\subsection{The Number of Servers for \texorpdfstring{$\Gr$}{GASPr} with Precomputation}

For $\Gr$ we define $\Npre_r$ as the number of servers required, $ \Npre_r = |(\alpha_I + \beta_I) \cup (\alpha_I + \beta_R) \cup (\alpha_R + \beta_I)|$ where each of these sets are as defined in the previous subsection. 
The following theorem establishes this number explicitly.

\begin{thm}\label{thm:GasprServers1}
    Consider the polynomial code $\Gr$ in the precomputation setting, with parameters $K$, $L$, $T$, and $r$.  Use Euclidean division to write $T = Ur + r_0$
    where $U = \left\lfloor T/r\right\rfloor$ and $0\leq r_0 < r$, as well as $ K + T - 1 = VK + K_0$ where $V = \left\lfloor(K+T-1)/K\right\rfloor$ and $0\leq K_0 < K$. Finally, let $W = L + U - 1$. Then, the number of servers $\Npre_r$ required by $\Gr$ in the precomputation setting is
    \begin{align*}
    \Npre_r &= KL + VK \\
    &+
    \left\{\begin{array}{ll}
        \max\{r,K_0\} + r(W-(V+1)) + r_0 & \text{if $V < W$} \\
        \max\{r_0, K_0\} & \text{if $V = W$} \\
        K_0 & \text{if $V > W$}
    \end{array}\right.
    \end{align*}
\end{thm}

For arbitrary $K$ and $L$, the expression for $\Npre_r$ in Theorem~\ref{thm:GasprServers1} need not be symmetric in $K$ and $L$.  However, by transposing $A$ and $B$ we are free to switch between $K$ and $L$.  So, if $\Npre_r(K,L)$ denotes the expression in Theorem~\ref{thm:GasprServers1} as a function of $K$ and $L$, the user only requires the amount of servers to be $\widehat{N}_r^{\mathrm{pre}} = \min\{\Npre_r(K,L), \Npre_r(L,K)\}$.

In the rest of this section, we examine the codes $\Gsmall$ and $\Gbig$ within the precomputation setting. These are specific instances of the code $\Gr$, so we can apply Theorem~\ref{thm:GasprServers1} to determine the number of servers for each case.

\subsection{\texorpdfstring{$\Gsmall$}{GASPsmall} with Precomputation}

The polynomial code $\Gsmall$ was introduced in~\cite{GASP/DOliveiraRK20}, and is equivalent to the code $\Gr$ with $r=1$. Writing $\Npre_{\mathrm{small}} = \Npre_1$ to be the number of servers required, we can use Theorem~\ref{thm:GasprServers1} with $r = 1$ and find
\begin{equation}\label{Eq:GsmallNpre}
    \Npre_{\mathrm{small}} = KL + K + L + 2T - 4 - \left\lfloor \frac{T-2}{K}\right\rfloor
\end{equation}
which follows from a straightforward computation. Symmetrizing with respect to $K$ and $L$ as in the discussion following Theorem~\ref{thm:GasprServers1}, the user can reduce this number to
\begin{equation}\label{Eq:GsmallNpre_symm}
\widehat{N}_{\mathrm{small}}^{\mathrm{pre}} = KL + K + L + 2T - 4 - \left\lfloor \frac{T-2}{m}\right\rfloor
\end{equation}
servers, where $m = \min\{K,L\}$.

\subsection{\texorpdfstring{$\Gbig$}{GASPbig} with Precomputation}

The polynomial code $\Gbig$ is equivalent to $\Gr$ with $r = \min\{K,T\}$, and was introduced in~\cite{GASP/DOliveiraRK20} along with $\Gsmall$. We can again appeal to Theorem~\ref{thm:GasprServers1} to compute the value $\Npre_{\mathrm{big}} = \Npre_{\min\{K,T\}}$ which gives the required number of servers for this scheme. We have
\begin{equation}\label{Eq:GbigNpre}
    \Npre_{\mathrm{big}} = \left\{
    \begin{array}{ll}
    2K + T - 1 & \text{if $L = 1$} \\
    KL + LT + K - T & \text{if $2\leq L$, $T\leq K$} \\
    2KL - K + T & \text{if $2\leq L$, $K\leq T$}
    \end{array}\right.
\end{equation}
Again, the above result follows from a straightforward computation, by plugging $r = \min\{K,T\}$ into Theorem~\ref{thm:GasprServers1}.

Since the quantity $\Npre_{\mathrm{big}}$ is a conditional expression wherein the conditions depend on $K$ and $L$, it does not admit a simple uniform ``symmetrization'' $\widehat{N}_{\mathrm{big}}^{\mathrm{pre}}$ as did the analogous expression for $\Gsmall$ with precomputation.

However, we can symmetrize the expression for $\Npre_{\mathrm{big}}$ on a case-by-case basis, which will suffice for comparison of $\Gsmall$ and $\Gbig$. We obtain
\begin{equation}\label{Eq:GbigNpre_symm}
\widehat{N}_{\mathrm{big}}^{\mathrm{pre}} =
\left\{\begin{array}{ll}
2M + T - 1 & \text{if $m = 1$} \\
KL + mT + M - T & \text{if $2\leq m$, $T\leq m$} \\
2KL - M + T & \text{if $2\leq m$, $M\leq T$} \\
\end{array}\right.
\end{equation}
where $m = \min\{K,L\}$ and $M = \max\{K,L\}$.  Note, however, that the conditions listed in Equation~\eqref{Eq:GbigNpre_symm} are now no longer exhaustive.

\subsection{Comparing
\texorpdfstring{$\Gsmall$}{GASPsmall} and \texorpdfstring{$\Gbig$}{GASPbig} with Precomputation}

To determine which of $\Gsmall$ and $\Gbig$ has superior performance in various parameter regimes, we now compare the expressions $\widehat{N}_{\mathrm{small}}^{\mathrm{pre}}$ and $\widehat{N}_{\mathrm{big}}^{\mathrm{pre}}$ found in~\eqref{Eq:GsmallNpre_symm} and~\eqref{Eq:GbigNpre_symm}, respectively. We have the following result.

\begin{thm}\label{thm:comparison_Gbig_Gsmall_new}
    Let $m = \min\{K,L\}$ and $M = \max\{K,L\}$.  Then, the following statements hold.
    \begin{enumerate}
        \item If $m = 1$ or $T = 1$, then $\widehat{N}_{\mathrm{small}}^{\mathrm{pre}} = \widehat{N}_{\mathrm{big}}^{\mathrm{pre}}$.
        \item If $2\leq T\leq m$, then $\widehat{N}_{\mathrm{small}}^{\mathrm{pre}}\leq \widehat{N}_{\mathrm{big}}^{\mathrm{pre}}$, with equality if and only if $m = T = 2$.
        \item If $2\leq m$ and $\frac{Mm^2 - 2}{m-1} < T$, then $\widehat{N}_{\mathrm{big}}^{\mathrm{pre}} <  \widehat{N}_{\mathrm{small}}^{\mathrm{pre}}$.
    \end{enumerate}
\end{thm}

Theorem~\ref{thm:comparison_Gbig_Gsmall_new} cannot compare $\Gsmall$ and $\Gbig$ once $m < T \leq \frac{Mm^2-2}{m-1}$.
Where the two cross over, including the values of $T$ for which they tie, we discuss in Section~\ref{sec:comparison}. 
It is straightforward to see that no other $r$ can do better than the two endpoints: writing $K + T - 1 = VK + K_0$ as in Theorem~\ref{thm:GasprServers1}, the tally there can be rewritten as
\[
    N_r^{\text{pre}} = KL + (K+T-1) + \bigl[\, T + (L-1-V)\,r - \min\{r, K_0\} \,\bigr]_+ ,
\]
which is nondecreasing in $r$ when $V \leq L-2$, and nonincreasing otherwise. 
In either orientation of the degree table, the minimum over $1 \leq r \leq \min\{K,T\}$ is therefore attained at $r = 1$ or at $r = \min\{K,T\}$, and no intermediate chain length satisfies $N_r^{\text{pre}} < \min\{N_{\text{small}}^{\text{pre}}, N_{\text{big}}^{\text{pre}}\}$. 
This does not occur without precomputation, where $\Gr$ with $1 < r < \min\{K,T\}$ is strictly better than both endpoints for intermediate values of $T$~\cite{DegTable/DOliveiraRHK21}.
For $K = L = 4$, this takes place at $T \in \{4, 5, 6\}$, on the left side of Figure~\ref{fig:KL_fixed:T_grows}.

Ties, on the other hand, are possible. 
For $K = L = 4$ and $T \in \{9, 10\}$, all four codes $\mathsf{GASP}_1, \ldots, \mathsf{GASP}_4$ require the same number of servers ($37$ and $38$, respectively); for $T = 11$, the codes $\mathsf{GASP}_2$ and $\mathsf{GASP}_3$ tie with $\Gbig$ at $39$ servers, one fewer than $\Gsmall$. 
This can be seen on the right side of Figure~\ref{fig:KL_fixed:T_grows}.

\section{Bounds on the Number of Servers and Complexity Analysis}\label{sec:bounds}
\subsection{Bounds on the Number of Servers}\label{sec:BoundsOnServers}

\begin{thm}\label{thm:PrecompBound}
For any degree table in the precomputation setting, we have
\begin{align}\label{eq:bound_1}
KL+\max\{K,L\}+T-1 \le \Npre
.
\end{align}
If $2 \le \min\{K,L,T\}$, then the last bound can be improved %by one
to
\begin{align}\label{eq:bound_2}
KL+\max\{K,L\}+T \le \Npre
.
\end{align}
Also for any degree table with precomputation we have
\begin{align}\label{eq:bound_3}
KL+K+L+T+m-2-m\min\{K,L,T\} \le \Npre
,
\end{align}
for all $m \in \{1,\ldots,T\}$ and~\eqref{eq:bound_3} strongest for $m=1$.
Moreover, \eqref{eq:bound_3} for $m=1$ is always equal or stronger (if and only if $T < \min\{K,L\}$) than~\eqref{eq:bound_1}.
\end{thm}

With these bounds, we can show the optimality of $\Gsmall$ for certain cases.

\begin{thm}\label{thm:OptimalityGap}
If $K=1$ or $L=1$ or $T=1$, or if $T=2$ and $\min\{K,L\}=2$, then $\widehat{N}_{\mathrm{small}}^{\mathrm{pre}}$ achieves the bounds of Theorem~\ref{thm:PrecompBound}.

\end{thm}

Note that $\Gbig$ in the setting with no precomputation is such that $N_{\mathrm{big}} = 2KL +2T -1$. If $T = \delta N$, then $N_{\mathrm{big}} = \frac{2KL - 1}{1 - 2\delta}$. This means that $\delta \leq \frac{1}{2}$. Therefore, we cannot tolerate scenarios where the percentage of collusions is greater than $\frac{1}{2}$. In fact, this is true for any traditional linear SDMM scheme~\cite[Corollary~2]{DBLP:conf/itw/MakkonenH22}. However, with precomputation, we have $\Npre_{\mathrm{big}} = 2KL - M + T$. Then, $\Npre_{\mathrm{big}} = \frac{2KL - M}{1 - \delta}$. Thus, we can tolerate any fixed percentage (other than $1$) of the total number of servers. Therefore, in an SDMM scenario where we expect $60\%$ of servers to collude, precomputation is necessary.

\subsection{Time Complexity Analysis}\label{sec:complexity}

 In this section, we analyze the total time complexity of our schemes. The total time complexity includes communication and computation complexity components.  The communication complexity includes the encoding operations performed by the user, the computations performed by the server (in parallel), and the decoding time at the user. The communication complexity includes the time to send the data from the user to servers and back.

We show that precomputation reduces the order of the total time complexity when the number of colluding servers is a fixed percentage of the total number of servers. 

 We assume the following  setting:

    \begin{itemize}

        \item The matrices $A$ and $B$ are $n\times n$.

        \item The parameters are $K=L=n^\varepsilon$ where $\varepsilon \in [0,1]$.

        \item The computational complexity for the servers to multiply two $m\times m$ matrices is $\mathcal{O}(m^\omega)$.

        \item The time complexity of a computational operation in $\mathbb{F}_q$ is constant with respect to $n$,
        
        \item The time to transmit a symbol of $\mathbb{F}_q$ is constant with respect to $n$.

    \end{itemize}

We can then compute the total time complexity for $\Gbig$ without precomputation.

\begin{thm} \label{thm:complexity_no_precomp}
    Assume  %Setting~\ref{setting}, and 
    that the security parameter $T = \delta N$ where $\delta \in [0,\frac{1}{2})$. Then, the total time complexity for $\Gbig$ without precomputation is $\mathcal{O}(n^{\max\{\varepsilon+\omega-\varepsilon\omega,2+3\varepsilon\}})$.
\end{thm}

Setting $\varepsilon$ to the optimal value, we obtain the minimum time complexity for $\Gbig$.

\begin{cor}
The minimum total time complexity for $\Gbig$ in the setting of Theorem~\ref{thm:complexity_no_precomp} is $\mathcal{O}(n^{5-\frac{12}{\omega+2}})$ for $\varepsilon=\frac{\omega-2}{\omega+2}$.
\end{cor}

If the servers utilize the standard matrix multiplication algorithm with $\omega=3$, the total time complexity for $\Gbig$ to compute the product of two $n\times n$ matrices without precomputation is $\mathcal{O}(n^{2.6})$. This represents an increase in the order of the time complexity when compared to $\Gbig$ without precomputation but with a constant number of collusions, which was shown to be $\mathcal{O}(n^{2.5})$ in~\cite{rafael_note}.

Next, we establish the total time complexity of $\Gbig$ with precomputation.

\begin{thm} \label{thm:complexity_with_precomp}
    Assume %Setting~\ref{setting}, and 
    that the security parameter $T = \delta N$ where $\delta \in [0,1)$. Then, the total time complexity for $\Gbig$ with precomputation is $\mathcal{O}(n^{\max\{\varepsilon+\omega-\varepsilon\omega,2+2\varepsilon\}})$.
\end{thm}

Setting $\varepsilon$ to the optimal value, we obtain the minimum time complexity for $\Gbig$.

\begin{cor}
The minimum total time complexity for $\Gbig$ in the setting of Theorem~\ref{thm:complexity_with_precomp} is $\mathcal{O}(n^{4-\frac{6}{\omega+1}})$ for $\varepsilon=\frac{\omega-2}{\omega+1}$.
\end{cor}

Thus, precomputation allows for a reduction in the order of the total time complexity. For example, if the servers utilize the standard matrix multiplication algorithm with $\omega=3$, then the total time complexity for $\Gbig$ to compute the product of two $n\times n$ matrices with precomputation is $\mathcal{O}(n^{2.5})$, a reduction in order compared to $\mathcal{O}(n^{2.6})$ for the case without precomputation.

\section{Gap-Reduced Additive Secure Polynomial ($\Grasp$) Codes}\label{sec:GRASP}

Every polynomial code that we have considered thus far shares the information supports of $\Gr$, namely the intervals $\alpha_I = [K-1]$ and $\beta_I = K \cdot [L-1]$, and the codes differ only in the placement of their ``mask supports,'' $\alpha_R$ and $\beta_R$.
As we show in the next section, no placement of the mask supports, including those of the $\CAT$ and $\DOG$ constructions~\cite{CatDog}, $\SHIFT$ and $\COVER$~\cite{ShiftCover} codes, or the Cartesian and projective line designs~\cite{CartesianProjective} use fewer servers than the optimal $\Gr$ code.
With this in mind, we now introduce $\Grasp$, Gap-Reduced Additive Secure Polynomial codes, which enjoy simple mask supports and concentrate on the information supports instead.

We first note that decodability restricts the information supports by requiring the $KL$ desired degrees in $\alpha_I + \beta_I$ to be distinct, which holds if and only if the difference sets $\alpha_I - \alpha_I$ and $\beta_I - \beta_I$ intersect only at $\{0\}$ -- nothing requires the supports to be an interval or an arithmetic progression.
Secondly, if both mask supports are the same interval $c + [0:T-1]$ with $c$ greater than every desired degree, then their union is
\[
    (\alpha_I + \beta_R) \cup (\alpha_R + \beta_I) = c + \bigl( (\alpha_I \cup \beta_I) + [0:T-1] \bigr),
\]
whose size is $T$ plus the sum -- over the consecutive gaps of $\alpha_I \cup \beta_I$ -- of the smaller of $T$ and the gap.
When every gap is at most $T$, this union becomes a single interval, while if $\min (\alpha_I \cup \beta_I) = 0$, then
\begin{equation}\label{eq:grasp_span}
    \Npre = KL + T + s, \text{ for span } s = \max (\alpha_I \cup \beta_I).
\end{equation}
The number of servers then depends on the information supports only through their span $s$.
The code $\Gbig$ is the special case in which $\alpha_I = [K-1]$ is an interval: this forces every nonzero difference of $\beta_I$ to be at least $K$, hence $s \geq K(L-1)$, and~\eqref{eq:grasp_span} becomes $2KL - K + T$ of~\eqref{Eq:GbigNpre} for all $T \geq K$.
Any pair of information supports of smaller span will then save $K(L-1) - s$ servers for every $T$ that is at least the size of its largest gap.

\begin{defn}[$\Grasp$ codes]\label{def:GRASP}
Let $K,L \geq 2$ and $T \geq 1$, and write $\alpha = (\cA \mid \cR)$ and
$\beta = (\cB \mid \cS)$ in~\eqref{eq:f_g_polys}.
\begin{enumerate}
    \item \emph{Common interval:} Let $P$ and $Q$ be sets of nonnegative
    integers with $|P| = K$, $|Q| = L$, $\min\{P\} = \min\{Q\} = 0$, and $|P+Q| = KL$,
    and let $c > \max\{P\} + \max\{Q\}$.
    Then
    \[
        \cA=P, \quad \cB=Q, \quad \cR = \cS = c + [0:T-1].
    \]

    \item \emph{Split:} Put $a = \lceil K/2 \rceil$ and $b = K - a$, and take
    \begin{align*}
        \cA &= [0:a-1] \cup \bigl( aL + [0:b-1] \bigr), \\
        \cB &= a \cdot [0:L-1], \\
        \cR &= \cS = 2aL + [0:T-1],
    \end{align*}
    in whichever orientation of $(K,L)$ gives fewer servers.
    This is the common-interval code for $P = [0:a-1] \cup (aL + [0:b-1])$ and $Q = a \cdot [0:L-1]$, whose union has span $aL + b - 1$ and largest gap $a$.

    \item \emph{Unified:} Choose $\lceil K/2 \rceil \leq a \leq K$ and
    $1 \leq r \leq \min\{a, T\}$, put $b=K-a$ and $c=2aL$, and take
    \begin{align*}
        \cA &= [0:a-1] \cup \bigl( aL + [0:b-1] \bigr), \\
        \cB &= a \cdot [0:L-1], \\
        \cR &= c + \gap(T; a,r), \quad \cS = c + [0:T-1],
    \end{align*}
    where $[0:-1] = \varnothing$ for $a=K$, and $\gap(T; a,r) = \{a \lfloor j/r \rfloor + (j \bmod r) \mid 0 \leq j < T\}$ takes $r$ consecutive integers from each block of length $a$, exactly as $\alpha_R$ does for $\Gr$.
    The choice $a = K$ is $\Gr$ with both mask supports shifted by $KL$, and the choice $a = \lceil K/2 \rceil$, $r = \min\{a, T\}$ is the split version.
\end{enumerate}
\end{defn}

Every member is decodable and $T$-secure over a sufficiently large field.
In the split and unified codes, the nonzero differences of $\cA$ lie in $[1:a-1] \cup [a(L-1) + 1 : aL + b - 1]$, while those of $\cB$ are the multiples of $a$ in $[a:a(L-1)]$, hence the desired degrees are distinct.
In the common interval code, this corresponds to $|P+Q| = KL$.
The shift $c$ places every mixed degree above every desired degree, the random exponents are pairwise distinct, and the number of servers is $KL$ plus the number of distinct mixed degrees, exactly as for $\Gr$.

The unified code interpolates between $\Gr$ and the split code in two ways.
The parameter $a$ shifts $b=K-a$ information exponents out of the interval $[0:a-1]$ into a second block placed just beyond $a \cdot [0:L-1]$, which lets $\beta_I$ be dilated by $a$ rather than by $K$ and roughly halves the span of $\cA \cup \cB$.
The parameter $r$ reintroduces gaps into one mask support.
The split code uses
\begin{equation}\label{eq:split_count}
    \begin{split}
        \Npre_{\mathrm{split}} = KL + T + \min \bigl\{ & L - 1 + (K - 1) \min\{T, \lceil L/2 \rceil\},\\
        & K - 1 + (L - 1) \min \{T, \lceil K/2 \rceil \}\bigr\}
    \end{split}
\end{equation}
servers.
For $K = L = n \geq 4$, this is fewer than the best $\Gr$ code once $T \geq n^2/2 - n + 3$ for even $n$, or $T \geq n(n-1)/2 + 3$ for odd $n$.
For $T \geq (n-1)^2$, this saves $(n-1)(n-2)/2$ servers when $n$ is even, and $(n-1)(n-3)/2$ when $n$ is odd, which grows quadratically.
There is no improvement in the $m=2$ case, consistent with the bound~\eqref{eq:bound_2}, which $\Gbig$ attains.
The gaps in the mask support produce this improvement for moderate $T$, where interval supports cannot compete.
Theorem~\ref{thm:grasp4L4} below is the instance with $a=2$ and $r=1$ at $K=T=4$: it represents an explicit family of $(4,L,4)$ degree tables with $5L+7$ servers, one fewer than the best $\Gr$ code for every $L \geq 3$, and a parameter-specific converse of $5L+5$ shows that these tables are within two servers of optimal.
The remainder of this section expounds upon this instance; the counts of the split and unified codes, and the comparison with the codes of the literature, are taken up in the next section.

\subsection{Comparison with \texorpdfstring{$\GASP_r$}{GASPr}}

\begin{thm}\label{thm:grasp4L4}
For every $L \geq 2$, define a $(4,L,4)$ degree table by
\[
\begin{aligned}
    A &= \{0,1,2L,2L+1\}, \quad B = 2[0:L-1],\\
    R &= 4L+2[0:3], \quad  S = 4L+[0:3].
\end{aligned}
\]
Equivalently,
\[
\begin{aligned}
    \alpha &= \bigl(0,1,2L,2L+1\mid 4L,4L+2,4L+4,4L+6\bigr), \\
    \beta &= \bigl(0,2,\ldots,2(L-1)\mid 4L,4L+1,4L+2,4L+3\bigr).
\end{aligned}
\]
This degree table is decodable and $4$-secure, with $\Npre_{L}=5L+7$.
\end{thm}

At $L=4$, the construction becomes
\begin{align*}
    \alpha &= (0,1,8,9\mid16,18,20,22), \\
    \beta &= (0,2,4,6\mid16,17,18,19),
\end{align*}
with $\Npre=27$.
At $L=3$, transposition gives
\begin{align*}
    \alpha &= (0,2,4\mid12,13,14,15), \\
    \beta &= (0,1,6,7\mid12,14,16,18),
\end{align*}
with $\Npre=22$.
We present the degree tables in Figure~\ref{tab:GRASP_examples2}, and highlight that these schemes are in some sense tailored for precomputation, and perform worse without.
This peculiarity highlights that the best code before precomputation may not remain the best after precomputation, and can be outperformed by code that were previously worse.

Substitution into the precomputation formula for $\GASP_r$ gives, for $L\geq 3$,
\[
    \begin{array}{c|cccc}
        (K,L,T) & r=1 & r=2 & r=3 & r=4 \\ \hline
        (4,L,4) & 5L+8 & 6L+5 & 7L+2 & 8L
    \end{array}.
\]
For the transposed orientation and $L\geq 4$,
\[
    \begin{array}{c|cccc}
    (K,L,T) & r=1 & r=2 & r=3 & r=4 \\ \hline
    (L,4,4) & 5L+8 & 5L+9 & 5L+10 & 5L+12
    \end{array}.
\]
For $r\geq T=4$ the first four mask degrees form a single chain, so every $r>4$ in the transposed orientation gives the same table as $r=4$.
When $L=3$, the first orientation gives $(23,23,23,24)$ and the transposed orientation gives $(23,24,25)$.
Consequently, for $L\geq 3$, we have that
\[
    \min_{\substack{\text{orientation} \\ r}} \Npre_r = 5L+8,
\]
and therefore for $L\geq 3$
\[
    \Npre_{\Grasp} = 5L+7 = \Npre_{\text{best }\Gr}-1.
\]
Thus once $L \geq 3$, this family performs better than the entire $\Gr$ family, including $\Gsmall$, $\Gbig$, every intermediate $r$, and both orientations.
For $L=2$, this family enjoys a threshold of $17$ while the best $\GASP_r$ threshold is $16$, so it performs worse.

\begin{figure*}[!t]
 \includegraphics[trim={2cm 4.9in 1cm 4.2in}, clip, width=\textwidth]{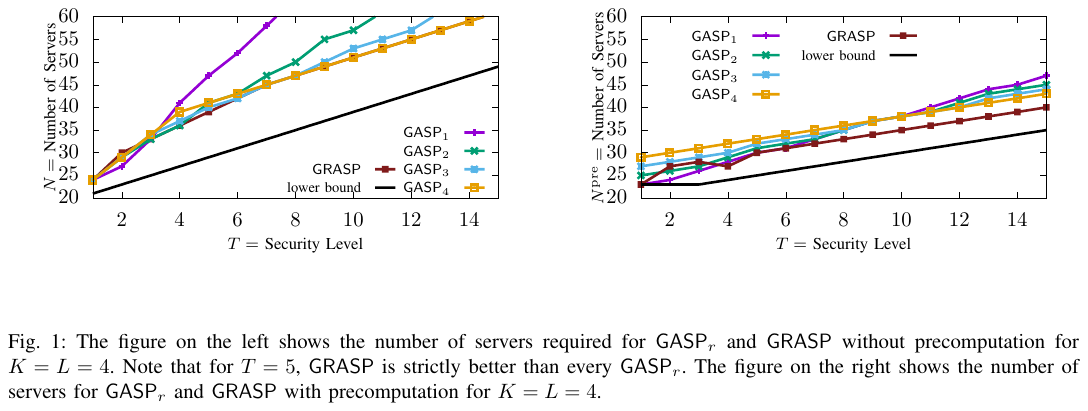} 
  \vspace*{-3mm}
  \caption{The figure on the left shows the number of servers required for $\Gr$ and $\Grasp$ without precomputation for $K=L=4$. Note that for $T=5$, $\Grasp$ is strictly better than every $\Gr$. The figure on the right shows the number of servers for $\Gr$ and $\Grasp$ with precomputation for $K=L=4$.
   }
  \label{fig:KL_fixed:T_grows2}
\end{figure*}

We demonstrate the performance of $\Grasp$ in Figure~\ref{fig:KL_fixed:T_grows2}, which showcases that (excepting $K=L=4$ and $T=5$) $\Gr$ performs better without precomputation, but in the precomputation model $\Grasp$ generally requires fewer servers, and is asymptotically only $2$ servers from optimal.
In Figure~\ref{tab:GRASP_examples2}, we present the degree tables for the two specific examples presented above, which require fewer servers than the best $\Gr$ using precomputation.

\subsection{Optimality}

The general converse from Equation~\eqref{eq:bound_2} gives
\[
    \Npre\geq KL+\max(K,L)+T
\]
once $K,L,T\geq 2$, so $\Npre \geq 20$ for $L=3$ and $\Npre \geq 5L+4$ for $L\geq 4$.
A parameter-specific argument turns this into the following:

\begin{thm}\label{thm:4L4Npre}
For every admissible $(4,L,4)$ degree table with $L\geq 3$,
\[
    \Npre\geq 5L+5.
\]
\end{thm}

The argument uses only the uniqueness of the $4L$ desired degrees and their disjointness from $C\cup E$, so it holds whether or not $A+B$ is also required to be distinct $R+S$.

Combining the converse with our construction implies that for $L\geq 3$, we have
\[
    5L + 5 \leq N^*(4,L,4) \leq 5L+7.
\]
Thus, this family is asymptotically optimal, and is always within two servers of optimal.

Note that exact optimality would follow from the stronger inequality
\[
    \left|(R+B) \cup(A+S) \right| \geq L+7.
\]

\begin{figure*}[!t]
\setlength{\tabcolsep}{0.33em}
\begin{subtable}{0.24\textwidth}
\resizebox{\columnwidth}{!}{\begin{tabular}{c|cccc|cccc}
  & 0& 1& 6& 7&\cellcolor{blue!25}12&\cellcolor{blue!25}14&\cellcolor{blue!25}16&\cellcolor{blue!25}18\\
\toprule
0& \cellcolor{red!25}0& \cellcolor{red!25}1& \cellcolor{red!25}6& \cellcolor{red!25}7& \cellcolor{gray!25}12& \cellcolor{gray!25}14& \cellcolor{gray!25}16& \cellcolor{gray!25}18\\
2& \cellcolor{red!25}2& \cellcolor{red!25}3& \cellcolor{red!25}8& \cellcolor{red!25}9& 14& 16& 18& \cellcolor{gray!25}20\\
4& \cellcolor{red!25}4& \cellcolor{red!25}5& \cellcolor{red!25}10& \cellcolor{red!25}11& 16& 18& 20& \cellcolor{gray!25}22\\
\midrule
\cellcolor{green!25}12& 12& \cellcolor{gray!25}13& 18& \cellcolor{gray!25}19& \cellcolor{gray!25}24& \cellcolor{gray!25}26& \cellcolor{gray!25}28& \cellcolor{gray!25}30\\
\cellcolor{green!25}13& 13& 14& 19& 20& \cellcolor{gray!25}25& \cellcolor{gray!25}27& \cellcolor{gray!25}29& \cellcolor{gray!25}31\\
\cellcolor{green!25}14& 14& \cellcolor{gray!25}15& 20& \cellcolor{gray!25}21& 26& 28& 30& \cellcolor{gray!25}32\\
\cellcolor{green!25}15& 15& 16& 21& 22& 27& 29& 31& \cellcolor{gray!25}33\\
\bottomrule
\end{tabular}}
\caption{$(3,4,4)$, $N\!=\!32$}\label{tab:GRASP344.2}
\end{subtable}
\begin{subtable}{0.24\textwidth}
\resizebox{\columnwidth}{!}{\begin{tabular}{c|cccc|cccc}
  & 0& 2& 4& 6&\cellcolor{blue!25}16&\cellcolor{blue!25}17&\cellcolor{blue!25}18&\cellcolor{blue!25}19\\
\toprule
0& \cellcolor{red!25}0& \cellcolor{red!25}2& \cellcolor{red!25}4& \cellcolor{red!25}6& \cellcolor{gray!25}16& \cellcolor{gray!25}17& \cellcolor{gray!25}18& \cellcolor{gray!25}19\\
1& \cellcolor{red!25}1& \cellcolor{red!25}3& \cellcolor{red!25}5& \cellcolor{red!25}7& 17& 18& 19& \cellcolor{gray!25}20\\
8& \cellcolor{red!25}8& \cellcolor{red!25}10& \cellcolor{red!25}12& \cellcolor{red!25}14& \cellcolor{gray!25}24& \cellcolor{gray!25}25& \cellcolor{gray!25}26& \cellcolor{gray!25}27\\
9& \cellcolor{red!25}9& \cellcolor{red!25}11& \cellcolor{red!25}13& \cellcolor{red!25}15& 25& 26& 27& \cellcolor{gray!25}28\\
\midrule
\cellcolor{green!25}16& 16& 18& 20& \cellcolor{gray!25}22& \cellcolor{gray!25}32& \cellcolor{gray!25}33& \cellcolor{gray!25}34& \cellcolor{gray!25}35\\
\cellcolor{green!25}18& 18& 20& 22& 24& 34& 35& \cellcolor{gray!25}36& \cellcolor{gray!25}37\\
\cellcolor{green!25}20& 20& 22& 24& 26& 36& 37& \cellcolor{gray!25}38& \cellcolor{gray!25}39\\
\cellcolor{green!25}22& 22& 24& 26& 28& 38& 39& \cellcolor{gray!25}40& \cellcolor{gray!25}41\\
\bottomrule
\end{tabular}}
\caption{$(4,4,4)$, $N\!=\!37$}\label{tab:GRASP444.2}
\end{subtable}
\begin{subtable}{0.24\textwidth}
\resizebox{\columnwidth}{!}{\begin{tabular}{c|cccc|cccc}
  & 0& 1& 6& 7&\cellcolor{blue!25}12&\cellcolor{blue!25}14&\cellcolor{blue!25}16&\cellcolor{blue!25}18\\
\toprule
0& \cellcolor{red!25}0& \cellcolor{red!25}1& \cellcolor{red!25}6& \cellcolor{red!25}7& \cellcolor{gray!25}12& \cellcolor{gray!25}14& \cellcolor{gray!25}16& \cellcolor{gray!25}18\\
2& \cellcolor{red!25}2& \cellcolor{red!25}3& \cellcolor{red!25}8& \cellcolor{red!25}9& 14& 16& 18& \cellcolor{gray!25}20\\
4& \cellcolor{red!25}4& \cellcolor{red!25}5& \cellcolor{red!25}10& \cellcolor{red!25}11& 16& 18& 20& \cellcolor{gray!25}22\\
\midrule
\cellcolor{green!25}12& 12& \cellcolor{gray!25}13& 18& \cellcolor{gray!25}19&\\
\cellcolor{green!25}13& 13& 14& 19& 20&\\
\cellcolor{green!25}14& 14& \cellcolor{gray!25}15& 20& \cellcolor{gray!25}21&\\
\cellcolor{green!25}15& 15& 16& 21& 22&\\
\bottomrule
\end{tabular}}
\caption{$(3,4,4)$, $\Npre\!=\!22$}\label{tab:GRASP344-pre.2}
\end{subtable}
\begin{subtable}{0.24\textwidth}
\resizebox{\columnwidth}{!}{\begin{tabular}{c|cccc|cccc}
  & 0& 2& 4& 6&\cellcolor{blue!25}16&\cellcolor{blue!25}17&\cellcolor{blue!25}18&\cellcolor{blue!25}19\\
\toprule
0& \cellcolor{red!25}0& \cellcolor{red!25}2& \cellcolor{red!25}4& \cellcolor{red!25}6& \cellcolor{gray!25}16& \cellcolor{gray!25}17& \cellcolor{gray!25}18& \cellcolor{gray!25}19\\
1& \cellcolor{red!25}1& \cellcolor{red!25}3& \cellcolor{red!25}5& \cellcolor{red!25}7& 17& 18& 19& \cellcolor{gray!25}20\\
8& \cellcolor{red!25}8& \cellcolor{red!25}10& \cellcolor{red!25}12& \cellcolor{red!25}14& \cellcolor{gray!25}24& \cellcolor{gray!25}25& \cellcolor{gray!25}26& \cellcolor{gray!25}27\\
9& \cellcolor{red!25}9& \cellcolor{red!25}11& \cellcolor{red!25}13& \cellcolor{red!25}15& 25& 26& 27& \cellcolor{gray!25}28\\
\midrule
\cellcolor{green!25}16& 16& 18& 20& \cellcolor{gray!25}22&\\
\cellcolor{green!25}18& 18& 20& 22& 24&\\
\cellcolor{green!25}20& 20& 22& 24& 26&\\
\cellcolor{green!25}22& 22& 24& 26& 28&\\
\bottomrule
\end{tabular}}
\caption{$(4,4,4)$, $\Npre\!=\!27$}\label{tab:GRASP444-pre.2}
\end{subtable}
\caption{The degree tables for the two $\mathsf{GRASP}$ examples $(3,4,4)$ and $(4,4,4)$. The gray regions represent the first time a number appears in a degree table, and denotes the number of distinct nodes needed.
Without precomputation (Tables~\ref{tab:GRASP344.2},~\ref{tab:GRASP444.2}) the mask-mask block is counted; with precomputation (Tables~\ref{tab:GRASP344-pre.2},~\ref{tab:GRASP444-pre.2}) it is computed offline and omitted, giving $\Npre=22$ and $\Npre=27$, one server fewer than the best $\Gr$ code in each case.
 }
\label{tab:GRASP_examples2}
\end{figure*}

\subsection{Difference-Disjoint Sets for Large \texorpdfstring{$T$}{T}}
\label{sec:span}

The $\Grasp$ codes of Section~\ref{sec:GRASP} replace the information supports of $\Gr$ with any pair of sets with non-intersecting pairwise sums.
This section considers the asymptotic case where $T$ is large.
When both random supports are the same interval, the number of servers is $KL+T+s$, where $s$ is the span of the union of the two information supports (Theorem~\ref{thm:span}).
The dependence on $T$ is then linear with unit slope, and the problem no longer relies on $T$: it is the packing problem of finding two difference-disjoint sets of sizes $K$ and $L$ of least span, which we denote $s^*(K,L)$.
We bound $s^*$ above and below (Lemma~\ref{lem:span_bounds}), determine it exactly for $K=4$ (Theorem~\ref{thm:span_four}) -- which shows that the information pair of Theorem~\ref{thm:grasp4L4} is the best possible for this construction -- and compare with $\Gr$ (Theorem~\ref{thm:span_vs_gasp}).
We find that for every fixed $K$ and $L$, the optimal number of servers satisfies
\[
    KL + M + T \leq \Nint(K,L,T) \leq KL + s^*(K,L) + T,
\]
a window which does not depend on $T$.
The consequence is that $\Grasp$ requires strictly fewer servers than every $\Gr$ code for large $T$ once $m \geq 4$ or when $m=3$ and $M$ is even, by a margin that grows quadratically in the partition parameters.
Here $\Nint(K,L,T)$ denotes the least $\Npre$ over all decodable and $T$-secure integer degree tables.

For a nonempty finite set $X$ of integers write $\widehat X = X -\min\{X\}$, and define
\[
 s^*(K,L) = \min \max\bigl\{ \widehat P \cup \widehat Q \bigr\}
\]
to be the minimum over all pairs of finite sets of integers with $|P|=K$, $|Q|=L$, and $(P-P) \cap (Q-Q) = \{0\}$.
This difference condition is equivalent to the map $(p, q) \mapsto p + q$ being injective on $P \times Q$.
% , that is, to $|P+Q| = KL$.

\begin{lem}\label{lem:span_bounds}
    For all $K,L \geq 1$,
    \[
        \max \Bigl\{ K+L-2, \Bigl\lceil \frac{KL-1}{2} \Bigr\rceil \Bigr\} \leq s^*(K,L) \leq \sigma(K,L),
    \]
    where
    \[
        \sigma(K,L) = \min \Bigl\{ (K-1) \Bigl\lceil \frac{L}{2} \Bigr\rceil + L - 1, (L-1) \Bigl\lceil \frac{K}{2} \Bigr\rceil + K - 1 \Bigr\}.
    \]
\end{lem}

\begin{thm}\label{thm:span}
    Let $P,Q$ be sets of integers with $|P|=K$, $|Q|=L$, and $(P-P) \cap (Q-Q) = \{0\}$, translated so that $P,Q \subseteq [0:s]$ and $0 \in P \cap Q$, where $s = \max\{ \widehat P \cup \widehat Q \}$.
    For any $T \geq 1$ and any integer $c > 2s$, the degree table
    \[
        \alpha_I = P, \quad \beta_I = Q, \quad \alpha_R = \beta_R = c + [0:T-1]
    \]
    is decodable and $T$-secure, with
    \begin{align*}
        \Npre &= KL + W \leq KL + s + T, \\
        W &= \bigl| (P + [0:T-1]) \cup (Q + [0:T-1]) \bigr|,
    \end{align*}
    with equality $W = s + T$ whenever $T$ is at least the largest gap between consecutive elements of $P\cup Q$.
    From this, we get that
    \[
        \Nint(K,L,T) \leq KL + s^*(K,L) + T .
    \]
\end{thm}

The information supports of Theorem~\ref{thm:grasp4L4} are an application of this construction, but with the roles of $K$ and $L$ interchanged: for $P = \{0, 1, 2L, 2L + 1\}$ and $Q = 2 \cdot [0:L-1]$, we have $s = 2L + 1$, with largest gap $2$.
Hence for every $L \geq 2$ and $T \geq 2$, the common-interval choice of random degrees results in
\[
    \Npre = 6L + T + 1.
\]
This is a different code from the $(4,L,4)$ tables of Theorem~\ref{thm:grasp4L4}, which use the same information supports, but placed the random degrees specifically in order to attain $5L+7$ at $T=4$; the two agree in their information supports, but differ in the random ones.

The lower bound $s^*(4,L) \geq 2L$ of Lemma~\ref{lem:span_bounds} is one short of the span $2L+1$ of this pair.
To close this gap, we leverage a classical theorem by de Bruijn on the tiling of an interval.

\begin{lem}[from \cite{deBruijn,Nathanson}]\label{lem:deBruijn}
    Let $n \geq 1$ and let $X,Y$ be sets of nonnegative integers with $X \oplus Y = [0:n-1]$, meaning that every element of $[0:n-1]$ has a unique representation $x+y$ with $x \in X$ and $y \in Y$.
    Then there are integers $f_1, \ldots, f_t \geq 2$ with $f_1 \cdots f_t = n$ and a partition $[1:t] = I_X \cup I_Y$ such that, with $w_i = f_1 \cdots f_{i-1}$,
    \begin{align*}
        X &= \Bigl\{ \sum_{i \in I_X} d_i w_i \ \Big| \ 0 \leq d_i < f_i \Bigr\}, \\
        Y &= \Bigl\{ \sum_{i \in I_Y} d_i w_i \ \Big| \ 0 \leq d_i < f_i \Bigr\}.
    \end{align*}
    In particular $|X| = \prod\limits_{i \in I_X} f_i$ and
    $\max\{X\} = \sum\limits_{i \in I_X} w_i (f_i - 1)$, and likewise for $Y$.
\end{lem}

\begin{thm}\label{thm:span_four}
    It holds that $s^*(4,2) = 4$, and $s^*(4,L) = 2L + 1$ for every $L \geq 3$.
\end{thm}

Thus the pair $(4,L)$ is best possible using the span construction, and $6L + T + 1$ is the smallest number of servers that it can yield with $K = 4$.
More generally, since $s^*(K,L) \geq \lceil (KL-1)/2 \rceil$, the span construction alone cannot close the gap to the converse~\eqref{eq:bound_2}.
Theorem~\ref{thm:grasp4L4}, whose mixed support has size $L+7$ compared to $2L+5$ for the common interval at $T=4$, shows that structured random supports recover part of this difference at small $T$.

\begin{thm}\label{thm:span_vs_gasp}
    Fix $K,L \geq 2$.
    For all sufficiently large $T$, the best $\Gr$ code over all $r$ and both orientations uses $2KL - M + T$ servers.
    The span construction therefore matches or improves on every $\Gr$ code for all large $T$, and the split pair of Lemma~\ref{lem:span_bounds} is strictly better whenever $m \geq 4$, or $m=3$ and $M$ is even, by $KL - M - \sigma(K,L) \geq (M-1) \bigl( \lfloor m/2 \rfloor - 1 \bigr)$ servers.
\end{thm}

For the pair $(4,L)$, the large-$T$ count $6L + T + 1$ attains one fewer server over the best $\Gr$ code for $L=3$, and $L-1$ servers when $L \geq 4$.
In general, this construction has the optimal linear dependence on $T$, and restricts the optimum to the interval
\[
    KL + M + T \leq \Nint(K,L,T) \leq KL + s^*(K,L) + T
\]
for fixed $K$ and $L$, by~\eqref{eq:bound_2} and Theorem~\ref{thm:span}.
The width of this interval is $\sigma(K,L) - M$ at most, and by Lemma~\ref{lem:span_bounds}, its upper bound cannot be reduced below $KL + \lceil (KL-1)/2 \rceil + T$ by any choice of $P$ and $Q$.
We leave what happens between these cases for future work.

\section{Comparison with Other Schemes}\label{sec:comparison}

Consider a user who splits $A$ into $K$ row blocks and $B$ into $L$ column blocks, with masks of $T$ random blocks $R_t$, $S_t$, and wants to receive the $KL$ products $A_k B_\ell$ from servers who each compute $Z_i = X_i Y_i$, assured privacy against every set of at most $T$ colluding servers.
With precomputation, the user knows the mask-mask products $R_s S_t$ in advance, so can subtract their contribution from each response.
Throughout, set $m=\min\{K,L\}$ and $M=\max\{K,L\}$ as in Theorem~\ref{thm:comparison_Gbig_Gsmall_new}, and let $\gap(t;p,r) = \{p \lfloor j/r \rfloor + (j \bmod r) \mid 0 \leq j < t\}$ denote the first $t$ elements of $\bigcup\limits_{u \geq 0}(up + [0:r-1])$, so that $\alpha_R = KL + \gap(T;K,r)$ for $\Gr$.

For an integer degree table with information supports $\cA = \alpha_I, \cB = \beta_I$ and mask supports $\cR = \alpha_R, \cS = \beta_R$ of size $K,L,T,T$ respectively, the desired degrees are $D = \cA + \cB$ and the mixed degrees are $E = (\cA + \cS) \cup (\cR + \cB)$.
The mask-mask block $\cR + \cS$ is precomputed.
We say that a table is \emph{admissible} when $|D| = KL$ and $D \cap E = \varnothing$, in which case $\Npre = KL + |E|$ servers suffice over every sufficiently large field.
Shifting both mask supports by some $c > \max \{D\}$ causes $D \cap E = \varnothing$ automatically, so the least number of servers over all admissible integer degree tables is
\[
    \Nint(K,L,T) = KL + \min \bigl \{|(\cA + \cS) \cup (\cR + \cB)| \ \big| \  |\cA + \cB| = KL \bigr \},
\]
an integer packing problem.
Note that $|\cA + \cB| = KL$ is equivalent to $(\cA - \cA) \cap (\cB - \cB) = \{0\}$, which was the form we considered for $\Grasp$.

For $\Gr$ with $1 \leq r \leq \min\{K, T\}$, we take $\cA = [0:K-1]$, $\cB = K \cdot [0:L-1]$, $\cR = KL + \gap(T; K,r)$, and $\cS = KL + [0:T-1]$.
With $K + T - 1 = VK + K_0$ for $0 \leq K_0 < K$ as in Theorem~\ref{thm:GasprServers1},
\[
    \Npre_r = KL + (K + T - 1) + \bigl[ T + (L - 1 - V)r - \min\{r, K_0\} \bigr]_+,
\]
where $[x]_+ = \max\{x,0\}$, which is nondecreasing in $r$ if $L - 1 - V \geq 1$, and nonincreasing otherwise -- hence only the endpoints $r = 1$ ($\Gsmall$) and $r = \min\{K, T\}$ ($\Gbig$) need be considered.
Using the optimal orientation from Equation~\eqref{Eq:GsmallNpre_symm},
\begin{align*}
    \widehat{N}_{\mathrm{small}}^{\mathrm{pre}} &= KL + K + L + 2T - 4 - \left\lfloor \frac{T-2}{m}\right\rfloor, \\
    % \widehat N_{\mathrm{small}}^{\mathrm{pre}} &= KL + K + L + 2T - 3 - \Big\lceil \frac{T-1}{m} \Big\rceil, \\
    \widehat{N}_{\mathrm{big}}^{\mathrm{pre}} &=
    \begin{cases}
        2M + T - 1, & m = 1, \\
        KL + M + (m - 1)T, & m \geq 2, \ T \leq M, \\
        2KL - M + T, & m \geq 2, \ T \geq M.
    \end{cases}
\end{align*}

We can now define the benchmark against which we compare the other schemes: define $B = 2M + T - 1$ for $m = 1$, and otherwise
\[
    B = KL + \min \Bigl\{ K + L + 2T - 4 - \Bigl\lfloor \frac{T-2}{m} \Bigr\rfloor, \ KL - M + T \Bigr\},
\]
so that $B = \min\{\widehat{N}_{\mathrm{small}}^{\mathrm{pre}}, \widehat{N}_{\mathrm{big}}^{\mathrm{pre}}\}$ for every $T \geq 2$.

For $m \geq 2$, $\Gsmall$ is strictly better for $2 \leq T < t_1$ while $\Gbig$ outperforms once $T > t_2$, where
\[
    t_x = (m - 1)(M - 1) + 1 - \lfloor (M - x) / (m - 1) \rfloor,
\]
and the two codes tie for $t_1 \leq T \leq t_2$.
If $K = L = n$, then the ties occur at $T = (n-1)^2$ and $(n-1)^2 + 1$.

To compare with the state-of-the-art, we first note that $\DOG$, $\SHIFT$, and $\COVER$ never use fewer than $B$ servers once the mask-mask block is free -- but their performance derives from packing precisely that quadrant, which precomputation deletes.
$\CAT$ uses at least two more servers than $B$ over its range, and the shortened $\PG$ construction, in either orientation, is never better than $B$.
The schemes of~\cite{CartesianProjective} trade servers for field size: the Cartesian construction works over any field admitting an $[M+T,T]$ maximum distance separable (MDS) code, but always uses more servers than $B$, while the projective line construction achieves $B$ for $T=2$ over any field with $KL + K + L \mid q-1$.
Table~\ref{tab:comparison} summarizes these counts, and highlights that among all constructions in the literature we compare against, the best precomputation count is attained by a $\Gr$ endpoint ($\Gbig$ or $\Gsmall$).
The sole exception is $\Grasp$, which is never worse than $B$, and is strictly better in certain regimes.

\begin{table*}[!t]
\centering\small
\setlength{\tabcolsep}{5pt}\renewcommand{\arraystretch}{1.2}
\begin{tabular}{@{}>{\raggedright\arraybackslash}p{2.5cm}>{\raggedright\arraybackslash}p{6.5cm}>{\raggedright\arraybackslash}p{5.6cm}>{\raggedright\arraybackslash}p{2.5cm}@{}}\toprule
Family & Supports & $\Npre$ & Compare with $B$\\ \midrule
$\Gr$ & $\cA = [0:K-1]$, $\cB = K[0:L-1]$, $\cR = KL+\gap(T;K,r)$, $\cS = KL+[0:T-1]$ for $1 \leq r \leq \min\{K,T\}$ & Theorem~\ref{thm:GasprServers1} & $=B$ at an endpoint \\

$\DOG_{r,s}$~\cite{CatDog}, $\SHIFT_{r,s}$ and $\COVER_r$~\cite{ShiftCover} & Gap supports with period $p=K+r$ & See citations. Interval does not benefit from larger $p$ & $\geq B$ \\

$\CAT$~\cite{CatDog} & Cyclic supports in $\mathbb{Z}/n\mathbb{Z}$ with $n = K_* L_* + (T-1)^2$, where $K_* \geq K+1$ and $L_* \geq L+1$ are the least integers coprime to $T-1$. Requires $K \geq L \geq T \geq 2$ and $n \mid q-1$ & $KL+K+L+2T-2$ & $\geq B+2$ \\

$\PG$~\cite{AGSDMM/MakkonenSH2023} (shortened, both orientations) & Functions on $y^2=f(x)$, $\deg(f)$ odd, over $\mathbb{F}_{s^2}$ with $s$ a large odd prime power & $KL+\min\{M+m T-1,\ \frac{KL + \varepsilon m}{2}+2T-1-\varepsilon\}$, $\varepsilon=1$ if $K,L$ odd and $0$ else & $\geq B$ \\

Cartesian~\cite{CartesianProjective} & Tensor grid, needs only an $[M+T,T]$ MDS code over $\mathbb{F}_q$ & $KL+T(K+L)$ & $>B$ for smallest fields \\

Projective line~\cite{CartesianProjective} & $T=2$, needs $KL+K+L\mid q-1$ & $KL+K+L$ & $=B$ \\ \midrule

$\Grasp$, common interval & $\cA=P$, $\cB=Q$ with $|P+Q|=KL$, $\cR=\cS=c+[0:T-1]$ & $KL + T + \sum\limits_j \min\{T,g_j\}$ for $g_j$ the gaps of $P\cup Q$. Equals $KL+T+s$ with $ s = \max\{P\cup Q\}$ once $T \geq \max\limits_j g_j$ & $<B$ for large $T$ iff $s<KL-M$. With $P,Q$ of least span, this holds for $m \geq 4$ and for $m=3$, $M$ even \\

$\Grasp$, split & $\cA=[0:a-1]\cup(aL+[0:b-1])$, $\cB=a[0:L-1]$, $a=\lceil K/2\rceil$, $b=K-a$, $\cR=\cS=2aL+[0:T-1]$ & $KL+T+\min\{M-1+(m-1)T,\ s\}$,
$s = \begin{cases}
        \tfrac{m M + m - 2}{2}, & m \text{ even}, \\
        \tfrac{(m + 1)M - 2}{2}, & m \text{ odd}, \ M \text{ even}, \\
        \tfrac{(m + 1)(M + 1) - 4}{2}, & \text{ both odd}.
    \end{cases}$
& $<B$ iff $T \geq T_{\mathrm{crit}}$ where $T_{\mathrm{crit}} = \tfrac{n^2}{2}-n+3$ ($n$ even), $\tfrac{n(n-1)}{2}+3$ ($n$ odd), for $K = L = n \geq4$. Never better for $m=2$ or $m=3$, $M$ odd \\

$\Grasp$, unified & $\cA=[0:a-1]\cup(aL+[0:b-1])$, $\cB=a[0:L-1]$, $\cR=c+\gap(T;a,r)$, $\cS=c+[0:T-1]$, $\lceil K/2\rceil\leq a\leq K$, $1\leq r\leq\min\{a,T\}$, $c=2aL$ & $a=K$ is $\Gr$, and $a=\lceil K/2\rceil$, $r=\min\{a,T\}$ is split & $\leq B$ always \\ \bottomrule
\end{tabular}
\caption{Precomputation counts of the constructions. $B$ is the best $\Gr$ code over all $r$ and both orientations. }\label{tab:comparison}
\label{tab:cat}
\end{table*}

\section{Conclusions}\label{sec:conclusion}

In this paper, we studied secure distributed matrix multiplication under the outer product partition wherein a user can precompute the products $R_t S_{t'}$ of the random matrices in advance.
In the degree table, these products occupy the lower-right block $\alpha_R + \beta_R$, so precomputation removes that portion from the count, hence the number of servers becomes $\Npre = |(\alpha_I + \beta_I) \cup (\alpha_I + \beta_R) \cup (\alpha_R + \beta_I)|$.
The design problem then asks to select degree tables which minimize $\Npre$ instead of $N$, but the two objectives can treat the same codes differently.
For example, taking $K = L = T = 4$, $\GASP_2$ is best before precomputation while $\GASP_1$ is best after applying it.

This change in the server count also changes the code design problem. A degree table that is effective without precomputation may obtain its savings from repeated degrees in the random–random block, while those repetitions provide no reduction in the online server count when the block is precomputed. Our results therefore show the value of designing schemes specifically for precomputation, rather than only applying precomputation to existing schemes. In particular, the $(4,L,4)$ $\Grasp$ family improves on every $\Gr$ code for $L\geq 3$, and the large-$T$ regime improve on $\Gr$ by a margin that increases quadratically in the partition parameters.

For the $\Gr$ family, we gave an exact formula for $\Npre_r$ (Theorem~\ref{thm:GasprServers1}), which is monotone in the chain length $r$, thus only the endpoints $\Gsmall$ and $\Gbig$ are consequential.
We compared these constructions (Theorem~\ref{thm:comparison_Gbig_Gsmall_new}), proved general lower bounds (Theorem~\ref{thm:PrecompBound}), and demonstrated that $\Gsmall$ attains them when $K=1$, $L=1$, $T=1$, or $T=2$ and $\min\{K,L\}=2$ (Theorem~\ref{thm:OptimalityGap}).

Without precomputation, every linear scheme requires $N \geq 2T + 1$, so at most half of the servers may collude; with precomputation however, $\Gbig$ needs $2KL - M + T$ servers, thus for any target fraction $\delta<1$ one can take $N=(2KL-M)/(1-\delta)$ servers to tolerate $\delta N$ colluders
In the regime $T = \delta N$, precomputation also lowers the order of the total time complexity (Theorems~\ref{thm:complexity_no_precomp} and~\ref{thm:complexity_with_precomp}), from $\mathcal{O}(n^{2.6})$ to $\mathcal{O}(n^{2.5})$ for $\omega=3$.

The best $\Gr$ code over all $r$ and both orientations is a natural benchmark, and Table~\ref{tab:cat} shows that none of the other constructions in the literature improve on it once the random--random block is free: $\DOG$, $\SHIFT$, and $\COVER$ obtain their performance by packing precisely that block -- which precomputation deletes -- while $\CAT$ and $\PG$ never reduce it, and the Cartesian and projective-line schemes trade servers for field size.
% All of these schemes retain the information supports of $\Gr$.
The $\Grasp$ codes of Section~\ref{sec:GRASP} instead alter the information supports, subject only to distinctness of the $KL$ desired degrees, and retain simple random supports.
The resulting $(4,L,4)$ family uses $5L + 7$ servers, one fewer than every $\Gr$ code for all $L \geq 3$ (Theorem~\ref{thm:grasp4L4}), and a parameter-specific converse of $5L + 5$ (Theorem~\ref{thm:4L4Npre}) shows that it is always within two servers of optimal.

For large $T$, with a common interval of random degrees, the number of servers is $KL+T+s$, where $s$ is the span of the union of the two information supports (Theorem~\ref{thm:span}).
This problem then reduces to a packing instance with no dependence on $T$: the least span $s^*(K,L)$ of two difference-disjoint sets.
We bounded $s^*$ (Lemma~\ref{lem:span_bounds}), determined it for $K=4$ (Theorem~\ref{thm:span_four}), and showed that the optimal value satisfies $KL+M+T \leq \Nint \leq KL+s^*(K,L)+T$, which does not depend on $T$.
In this regime, $\Grasp$ outperforms every $\Gr$ code by an amount that grows quadratically in the partition parameters whenever $m \geq 4$, or when $m=3$ and $M$ is even (Theorem~\ref{thm:span_vs_gasp}).

We identify a number of future directions.
First, in Theorem~\ref{thm:4L4Npre}, we note that $\Grasp$ would be perfectly optimal given the strengthened inequality,
\[
\left|(R+B)\cup(A+S)\right|\geq L+7,
\]
which we leave for future work.

Similarly, the precise value of $\Nint$ in the large-$T$ regime, the exact value of $s^*(K,L)$ beyond $K=4$, and whether the bounds $5L+5$ or $5L+6$ can be attained for some $(4,L,4)$ family are all open.

Finally, the user may elect to only precompute some part of the random--random block, which we term partial precomputation, so characterizing the trade-off between offline work and number of servers is a natural extension of this work.

\bibliographystyle{IEEEtran}
\bibliography{bib}

@inproceedings{Beaver,
  title={Efficient multiparty protocols using circuit randomization},
  author={Beaver, Donald},
  booktitle={Advances in Cryptology -- CRYPTO '91},
  pages={420--432},
  year={1992},
  organization={Springer}
}

@InProceedings{10.1007/978-3-030-36030-6_14,
author="Boyle, Elette
and Gilboa, Niv
and Ishai, Yuval",
editor="Hofheinz, Dennis
and Rosen, Alon",
title="Secure Computation with Preprocessing via Function Secret Sharing",
booktitle="Theory of Cryptography",
year="2019",
publisher="Springer International Publishing",
address="Cham",
pages="341--371",
isbn="978-3-030-36030-6"
}

@InProceedings{10.1007/978-3-642-32009-5_40,
author="Nielsen, Jesper Buus
and Nordholt, Peter Sebastian
and Orlandi, Claudio
and Burra, Sai Sheshank",
editor="Safavi-Naini, Reihaneh
and Canetti, Ran",
title="A New Approach to Practical Active-Secure Two-Party Computation",
booktitle="Advances in Cryptology -- CRYPTO 2012",
year="2012",
publisher="Springer Berlin Heidelberg",
address="Berlin, Heidelberg",
pages="681--700",
isbn="978-3-642-32009-5"
}

@InProceedings{10.1007/978-3-642-32009-5_38,
author="Damg{\aa}rd, Ivan
and Pastro, Valerio
and Smart, Nigel
and Zakarias, Sarah",
editor="Safavi-Naini, Reihaneh
and Canetti, Ran",
title="Multiparty Computation from Somewhat Homomorphic Encryption",
booktitle="Advances in Cryptology -- CRYPTO 2012",
year="2012",
publisher="Springer Berlin Heidelberg",
address="Berlin, Heidelberg",
pages="643--662",
isbn="978-3-642-32009-5"
}

@InProceedings{10.1007/978-3-319-39555-5_18,
author="Baum, Carsten
and Damg{\aa}rd, Ivan
and Toft, Tomas
and Zakarias, Rasmus",
editor="Manulis, Mark
and Sadeghi, Ahmad-Reza
and Schneider, Steve",
title="Better Preprocessing for Secure Multiparty Computation",
booktitle="Applied Cryptography and Network Security",
year="2016",
publisher="Springer International Publishing",
address="Cham",
pages="327--345",
isbn="978-3-319-39555-5"
}

@article{SDMM2/KakarES2019,
  author={Kakar, Jaber and Ebadifar, Seyedhamed and Sezgin, Aydin},
  journal={IEEE Access}, 
  title={On the Capacity and Straggler-Robustness of Distributed Secure Matrix Multiplication}, 
  year={2019},
  volume={7},
  number={},
  pages={45783--45799},
  doi={10.1109/ACCESS.2019.2908024}
}

@misc{AGSDMM/MakkonenSH2023,
      title={Algebraic Geometry Codes for Secure Distributed Matrix Multiplication}, 
      author={Okko Makkonen and Elif Saçıkara and Camilla Hollanti},
      year={2023},
      eprint={2303.15429},
      url={https://arxiv.org/abs/2303.15429},
      archivePrefix={arXiv},
      primaryClass={cs.IT}
}

@inproceedings{PC/YuMA2017,
author = {Yu, Qian and Maddah-Ali, Mohammad Ali and Avestimehr, A. Salman},
title = {Polynomial Codes: An Optimal Design for High-Dimensional Coded Matrix Multiplication},
year = {2017},
isbn = {9781510860964},
publisher = {Curran Associates Inc.},
address = {Red Hook, NY, USA},
booktitle = {Proceedings of the 31st International Conference on Neural Information Processing Systems},
pages = {4406--4416},
numpages = {11},
location = {Long Beach, California, USA},
series = {NIPS'17}
}

@article{SDMM1/TandonChang2018,
  author       = {Wei{-}Ting Chang and
                  Ravi Tandon},
  title        = {On the Capacity of Secure Distributed Matrix Multiplication},
  journal      = {CoRR},
  volume       = {abs/1806.00469},
  year         = {2018},
  url          = {http://arxiv.org/abs/1806.00469},
  eprinttype    = {arXiv},
  eprint       = {1806.00469},
  bibsource    = {dblp computer science bibliography, https://dblp.org}
}

@article{GASP/DOliveiraRK20,
  author       = {Rafael G. L. D'Oliveira and
                  El Rouayheb, Salim and
                  David A. Karpuk},
  title        = {{GASP} Codes for Secure Distributed Matrix Multiplication},
  journal      = {{IEEE} Trans. Inf. Theory},
  volume       = {66},
  number       = {7},
  pages        = {4038--4050},
  year         = {2020},
  url          = {https://doi.org/10.1109/TIT.2020.2975021},
  doi          = {10.1109/TIT.2020.2975021},
  bibsource    = {dblp computer science bibliography, https://dblp.org}
}

@article{DegTable/DOliveiraRHK21,
  author       = {Rafael G. L. D'Oliveira and
                  El Rouayheb, Salim and
                  Daniel Heinlein and
                  David A. Karpuk},
  title        = {Degree Tables for Secure Distributed Matrix Multiplication},
  journal      = {{IEEE} J. Sel. Areas Inf. Theory},
  volume       = {2},
  number       = {3},
  pages        = {907--918},
  year         = {2021},
  url          = {https://doi.org/10.1109/JSAIT.2021.3102882},
  doi          = {10.1109/JSAIT.2021.3102882},
  bibsource    = {dblp computer science bibliography, https://dblp.org}
}

@book{DBLP:books/daglib/0020358,
  author       = {Terence Tao and
                  Van H. Vu},
  title        = {Additive combinatorics},
  series       = {Cambridge studies in advanced mathematics},
  volume       = {105},
  publisher    = {Cambridge University Press},
  year         = {2007},
  isbn         = {978-0-521-85386-6},
  bibsource    = {dblp computer science bibliography, https://dblp.org}
}

@ARTICLE{jafar_batch2,
  author={Zhuqing Jia and Syed Ali Jafar} ,
  journal={IEEE Transactions on Information Theory}, 
  title={On the Capacity of Secure Distributed Batch Matrix Multiplication}, 
  year={2021},
  volume={67},
  number={11},
  pages={7420--7437},
  doi={10.1109/TIT.2021.3112952}}

@misc{karpuk2024modular,
      title={Modular Polynomial Codes for Secure and Robust Distributed Matrix Multiplication}, 
      author={David Karpuk and Razane Tajeddine},
      year={2024},
      eprint={2305.03465},
      url={https://arxiv.org/abs/2305.03465},
      archivePrefix={arXiv},
      primaryClass={cs.IT}
}

@ARTICLE{jafar_cross,
  author={Jia, Zhuqing and Sun, Hua and Jafar, {Syed Ali}},
  journal={IEEE Transactions on Information Theory}, 
  title={Cross Subspace Alignment and the Asymptotic Capacity of  {$X$}-Secure {$T$}-Private Information Retrieval}, 
  year={2019},
  volume={65},
  number={9},
  pages={5783--5798},
  doi={10.1109/TIT.2019.2916079}}

@ARTICLE{jafar_xsecure_tprivate,
  author={Jia, Zhuqing and Jafar, {Syed Ali}},
  journal={IEEE Transactions on Information Theory}, 
  title={{$X$}-Secure {$T$}-Private Information Retrieval From {MDS} Coded Storage With {B}yzantine and Unresponsive Servers}, 
  year={2020},
  volume={66},
  number={12},
  pages={7427--7438},
  doi={10.1109/TIT.2020.3013152}}

@misc{hermitian,
      title={{HerA} Scheme: Secure Distributed Matrix Multiplication via {H}ermitian Codes}, 
      author={Roberto A. Machado and Welington Santos and Gretchen L. Matthews},
      year={2023},
      eprint={2303.16366},
      url={https://arxiv.org/abs/2303.16366},
      archivePrefix={arXiv},
      primaryClass={cs.IT}
}

@INPROCEEDINGS{field_trace,
  author={Machado, Roberto Assis and D'Oliveira, Rafael G. L. and El Rouayheb, Salim and Heinlein, Daniel},
  booktitle={2021 XVII International Symposium ``Problems of Redundancy in Information and Control Systems'' (REDUNDANCY)}, 
  title={Field Trace Polynomial Codes for Secure Distributed Matrix Multiplication}, 
  year={2021},
  volume={},
  number={},
  pages={188--193},
  doi={10.1109/REDUNDANCY52534.2021.9606447}}

@ARTICLE{systematic_private,
  author={Zhu, Jinbao and Li, Songze},
  journal={IEEE Journal on Selected Areas in Information Theory}, 
  title={A Systematic Approach Towards Efficient Private Matrix Multiplication}, 
  year={2022},
  volume={3},
  number={2},
  pages={257--274},
  doi={10.1109/JSAIT.2022.3181144}}

@INPROCEEDINGS{rawad_latency,
  author={Bitar, Rawad and Parag, Parimal and El Rouayheb, Salim},
  booktitle={2017 IEEE International Symposium on Information Theory (ISIT)}, 
  title={Minimizing latency for secure distributed computing}, 
  year={2017},
  volume={},
  number={},
  pages={2900--2904},
  doi={10.1109/ISIT.2017.8007060}}

@INPROCEEDINGS{rafael_note,
  author={D'Oliveira, Rafael G. L. and El Rouayheb, Salim and Heinlein, Daniel and Karpuk, David},
  booktitle={2020 IEEE Conference on Communications and Network Security (CNS)}, 
  title={Notes on Communication and Computation in Secure Distributed Matrix Multiplication}, 
  year={2020},
  volume={},
  number={},
  pages={1--6},
  doi={10.1109/CNS48642.2020.9162296}}

@INPROCEEDINGS{upload_vs_download,
  author={Chang, Wei-Ting and Tandon, Ravi},
  booktitle={2019 IEEE Information Theory Workshop (ITW)}, 
  title={On the Upload versus Download Cost for Secure and Private Matrix Multiplication}, 
  year={2019},
  volume={},
  number={},
  pages={1--5},
  doi={10.1109/ITW44776.2019.8989342}}

@INPROCEEDINGS{bivariate,
  author={Hasircioglu, Burak and G\'omez-Vilardeb\'o, Jes\'us and G\"und\"uz, Deniz},
  booktitle={2021 IEEE International Symposium on Information Theory (ISIT)}, 
  title={Speeding Up Private Distributed Matrix Multiplication via Bivariate Polynomial Codes}, 
  year={2021},
  volume={},
  number={},
  pages={1853--1858},
  doi={10.1109/ISIT45174.2021.9517739}}

@ARTICLE{rawad_adaptive,
  author={Bitar, Rawad and Xhemrishi, Marvin and Wachter-Zeh, Antonia},
  journal={IEEE Transactions on Information Theory}, 
  title={Adaptive Private Distributed Matrix Multiplication}, 
  year={2022},
  volume={68},
  number={4},
  pages={2653--2673},
  doi={10.1109/TIT.2022.3143199}}

@ARTICLE{jafar_batch,
  author={Jia, Zhuqing and Jafar, {Syed Ali}},
  journal={IEEE Transactions on Information Theory}, 
  title={Cross Subspace Alignment Codes for Coded Distributed Batch Computation}, 
  year={2021},
  volume={67},
  number={5},
  pages={2821--2846},
  doi={10.1109/TIT.2021.3064827}}

@misc{kakar2020uplinkdownlink,
      title={Uplink-Downlink Tradeoff in Secure Distributed Matrix Multiplication}, 
      author={Jaber Kakar and Anton Khristoforov and Seyedhamed Ebadifar and Aydin Sezgin},
      year={2020},
      eprint={1910.13849},
      url={https://arxiv.org/abs/1910.13849},
      archivePrefix={arXiv},
      primaryClass={cs.IT}
}

@ARTICLE{inner_product,
  author={Mital, Nitish and Ling, Cong and G\"und\"uz, Deniz},
  journal={IEEE Transactions on Information Theory}, 
  title={Secure Distributed Matrix Computation With Discrete {F}ourier Transform}, 
  year={2022},
  volume={68},
  number={7},
  pages={4666--4680},
  doi={10.1109/TIT.2022.3158868}}

@INPROCEEDINGS{root_of_unity,
  author={Machado, Roberto Assis and Manganiello, Felice},
  booktitle={2022 IEEE Information Theory Workshop (ITW)}, 
  title={Root of Unity for Secure Distributed Matrix Multiplication: Grid Partition Case}, 
  year={2022},
  volume={},
  number={},
  pages={155--159},
  doi={10.1109/ITW54588.2022.9965858}}

@Article{oliver,
AUTHOR = {Byrne, Eimear and Gnilke, Oliver W. and Kliewer, J\"org},
TITLE = {Straggler- and Adversary-Tolerant Secure Distributed Matrix Multiplication Using Polynomial Codes},
JOURNAL = {Entropy},
VOLUME = {25},
YEAR = {2023},
NUMBER = {2},
ARTICLE-NUMBER = {266},
URL = {https://www.mdpi.com/1099-4300/25/2/266},
PubMedID = {36832632},
ISSN = {1099-4300},
DOI = {10.3390/e25020266}
}

@inproceedings{DBLP:conf/itw/MakkonenH22,
  author       = {Okko Makkonen and
                  Camilla Hollanti},
  title        = {General Framework for Linear Secure Distributed Matrix Multiplication
                  with {B}yzantine Servers},
  booktitle    = {{IEEE} Information Theory Workshop, {ITW} 2022, Mumbai, India, November
                  1-9, 2022},
  pages        = {143--148},
  publisher    = {{IEEE}},
  year         = {2022},
  url          = {https://doi.org/10.1109/ITW54588.2022.9965828},
  doi          = {10.1109/ITW54588.2022.9965828},
  bibsource    = {dblp computer science bibliography, https://dblp.org}
}

@inproceedings{kazemi2022degree,
  title={Degree Tables for Private Information Retrieval},
  author={Kazemi, Fatemeh and Wang, Ningze and D'Oliveira, Rafael G. L. and Sprintson, Alex},
  booktitle={2022 58th Annual Allerton Conference on Communication, Control, and Computing (Allerton)},
  pages={1--8},
  year={2022},
  organization={IEEE}
}

@INPROCEEDINGS{CatDog,
  author={Hofmeister, Christoph and Bitar, Rawad and Wachter-Zeh, Antonia},
  booktitle={2025 IEEE International Symposium on Information Theory (ISIT)}, 
  title={CAT and DOG: Improved Codes for Private Distributed Matrix Multiplication}, 
  year={2025},
  volume={},
  number={},
  pages={1--6},
  doi={10.1109/ISIT63088.2025.11195364}}

@misc{ShiftCover,
      title={Improved Degree Tables for Secure Distributed Matrix Multiplication}, 
      author={John Byrne and Rafael G. L. D'Oliveira and Michael Tait},
      year={2026},
      eprint={2609.06789},
      archivePrefix={arXiv},
      primaryClass={cs.IT},
      url={https://arxiv.org/abs/2609.06789}, 
}

@misc{CartesianProjective,
      title={Function Tables for Secure Distributed Matrix Multiplication}, 
      author={Rafael G. L. D'Oliveira and Giulia Gaggero and Arturo Jaramillo Gil and Hiram H. López and Cecilia Martínez-Reyes and Divyesh Vaghasiya},
      year={2026},
      eprint={2609.08154},
      archivePrefix={arXiv},
      primaryClass={cs.IT},
      url={https://arxiv.org/abs/2609.08154}, 
}

@misc{PoleGap,
      title={Algebraic Geometry Codes for Secure Distributed Matrix Multiplication}, 
      author={Okko Makkonen and Elif Saçıkara and Camilla Hollanti},
      year={2023},
      eprint={2303.15429},
      archivePrefix={arXiv},
      primaryClass={cs.IT},
      url={https://arxiv.org/abs/2303.15429}, 
}

@article{deBruijn,
  author  = {de Bruijn, N. G.},
  title   = {On number systems},
  journal = {Nieuw Archief voor Wiskunde},
  volume  = {4},
  pages   = {15--17},
  year    = {1956}
}

@misc{Nathanson,
      title={Additive systems and a theorem of de {B}ruijn}, 
      author={Melvyn B. Nathanson},
      year={2013},
      eprint={1301.6208},
      archivePrefix={arXiv},
      primaryClass={math.NT},
      url={https://arxiv.org/abs/1301.6208}, 
}

\clearpage

\appendix

\subsection{Sum Sets}
\begin{lem}[Sum Sets,~\cite{DBLP:books/daglib/0020358}]\label{lem:sumset}
    Let $A$ and $B$ be sets of integers. Then
$|A| +|B|-1 \leq |A+B|$. And if $2 \leq |A|, |B|$, then equality holds
if and only if $A$ and $B$ are arithmetic progressions with the
same common difference, i.e., $A = a+d[m]$ and $B = b+d[n]$
for $a, b, d,m, n \in \mathbb{Z}$.
\end{lem}

\subsection{Proof of Theorem~\ref{thm:GasprServers1}}\label{pf:GasprServers1}

\begin{proof}
We must compute an explicit formula for
\[
\Npre_r = |(\alpha_I+\beta_I) \cup (\alpha_I+\beta_R) \cup (\alpha_R+\beta_I)|
\]
for the polynomial code $\Gr$.  Clearly $\alpha_I + \beta_I = [KL - 1]$ and this set is disjoint from the union of the other two.  We therefore wish to calculate $\Npre_r - KL = |(\alpha_I+\beta_R)\cup (\alpha_R+\beta_I)|$.  To simplify computations it will be convenient to shift both sets by $KL$.  Given the definitions of these sets in the previous subsection, we have
\begin{equation}\label{alphaI_betaR}
    \alpha_I + \beta_R - KL = [K + T - 2]
\end{equation}

One also readily computes that
\begin{align}\label{alphaR_betaI}
\begin{split}
    \alpha_R &+ \beta_I - KL \\
    =& \bigcup_{u = 1}^U \left((u-1)K+[r-1]\right) \cup \left(UK + [r_0-1]\right)\\
    &+ K\cdot [L-1] \\
    =& \bigcup_{u = 1}^U\left((u-1)K + [r-1] + K\cdot [L-1]\right) \\
    &\cup \left(UK + [r_0-1] + K\cdot [L-1]\right) \\
    =& \bigcup_{w = 1}^W\left((w-1)K + [r-1]\right) \cup \left(WK + [r_0-1]\right)
\end{split}
\end{align}
with $U$ and $W$ as in the statement of the theorem.  Note that since $r\leq K$ this expresses $\alpha_R + \beta_I - KL$ as a union of \emph{disjoint} intervals, each contained in an interval of the form $J_w = [(w-1)K:wK-1]$, which are themselves disjoint and all of length $K$.

The sets $(w-1)K + [r-1]$ are the first $r$ integers in the intervals $J_w$.  By construction, the integer $V$ is the number of these intervals which are entirely contained in $[K + T - 2]$.  Thus we get a contribution of $VK$ to $\Npre_r$ from $J_w$ for $w = 1,\ldots,V$.

If $V < W$ and we set $w = V + 1$ in the final union expression in~\eqref{alphaR_betaI}, then $VK+[r-1]\subseteq J_{V+1}$ consists of the first $r$ integers of this latter interval.  Moreover, $[K+T-2]\cap J_{V+1} = VK + [K_0-1]$.  Thus the interval $J_{V+1}$ contributes $\max\{r,K_0\}$ to $\Npre_r$.  For $w = V+2,\ldots,W$, we get a contribution of $r$ from each of these remaining $W-(V+1)$ intervals $J_w$.  Finally, the interval $WK + [r_0-1]\subseteq J_{W+1}$ contributes an $r_0$, which finishes the computation for $V< W$.

The cases of $V = W$ and $V> W$ can be analyzed similarly, by calculating the contributions from each of the $J_w$.
\end{proof}

\subsection{Proof of Theorem~\ref{thm:comparison_Gbig_Gsmall_new}}\label{pf:comp}

Cases~1) and 2) follow from a straightforward comparison of the expressions for $\widehat{N}_{\mathrm{small}}^{\mathrm{pre}}$ and $\widehat{N}_{\mathrm{big}}^{\mathrm{pre}}$ found in Equations~\eqref{Eq:GsmallNpre_symm} and~\eqref{Eq:GbigNpre_symm}.  For Case~3), the assumption $\frac{Mm^2 - 2}{m-1} < T$ is equivalent to
\[
Mm < T - \frac{T-2}{m}.
\]
As $-2M - m + 4\leq 0$ we are free to add this quantity to the left-hand side, as well as replace $(T-2)/m$ with $\left\lfloor (T-2)/m\right\rfloor$.  Elementary algebra now provides us with the inequality
\[
2Mm - M + T < M + m + 2T - 4 - \left\lfloor \frac{T-2}{m}\right\rfloor
\]
which is equivalent to $\widehat{N}_{\mathrm{big}}^{\mathrm{pre}} < \widehat{N}_{\mathrm{small}}^{\mathrm{pre}}$, hence the result.

\subsection{Proof of Theorem~\ref{thm:PrecompBound}}\label{pf:PrecompBound}
\begin{proof}
To simplify the notation, we regard $\alpha_I$, $\alpha_R$, $\beta_I$, and $\beta_R$ as sets and vectors interchangeably, noticing that $N$ is invariant under reordering.
Abbreviate $A=\alpha_I+\beta_I$, $B=\alpha_I+\beta_R$, and $C=\alpha_R+\beta_I$.
Then, as $A \cap (B \cup C) = \emptyset$, $N=|A \cup B \cup C| = |A| + |B \cup C| = KL+|B \cup C|$.
Next, $|B \cup C| \ge |B| \ge |\alpha_I|+|\beta_R|-1 = K+T-1$ by Lemma~\ref{lem:sumset}.
Exchanging $B$ and $C$ in the last argument finishes the proof of~\eqref{eq:bound_1}.

For the second statement, assume $2 \le T$ and, without loss of generality, $2 \le L \le K$ and $N = KL+K+T-1$.

As $N = KL+K+T-1$, we have that $B \cup C = B$, i.e., $C \subseteq B$, and, by Lemma~\ref{lem:sumset}, $\alpha_I = a+d[K-1]$ and $\beta_R = b+d[T-1]$ for $a,b,d \in \mathbb{Z}$ and $1 \le d$.
Hence, $B=a+b+d[K+T-2]$ and, as $C \subseteq B$, $C \subseteq a+b+d\mathbb{Z}$.

Then, as $\alpha_R+\beta_I^{(1)} \subseteq C$ and $\alpha_R^{(1)}+\beta_I \subseteq C$, we have $\alpha_R = y+d\{m_1,\ldots,m_T\}$ and $\beta_I = x+d\{\lambda_1,\ldots,\lambda_L\}$ for some $x,y \in \mathbb{Z}$, $m_i \in \mathbb{Z}$ ($i=1,\ldots,T$), and $\lambda_i \in \mathbb{Z}$ ($i=1,\ldots,L$) with $m_1=\lambda_1=0$, $m_i < m_{i+1}$ ($i=1,\ldots,T-1$), and $\lambda_i < \lambda_{i+1}$ ($i=1,\ldots,L-1$).

Next, $A=\alpha_I+\beta_I=(a+d[K-1])+(x+d\{\lambda_1,\ldots,\lambda_L\})$ and all elements have to be distinct, hence $\lambda_i+K \le \lambda_{i+1}$ for all $i=1,\ldots,L-1$.
This implies $\lambda_0+(L-1)K \le \lambda_L$.

Finally, $C=\alpha_R+\beta_I=(y+d\{m_1,\ldots,m_T\})+(x+d\{\lambda_1,\ldots,\lambda_L\}) \subseteq B = a+b+d[K+T-2]$, hence $x+y=a+b+dv$ for a $v \in \mathbb{Z}$ with $0 \le v$ and $v+\{m_1,\ldots,m_T\}+\{\lambda_1,\ldots,\lambda_L\} \subseteq [K+T-2]$.
In particular, $v+m_T+\lambda_L \le K+T-2$ and as $T-1 \le m_T$, $(0)+(T-1)+((L-1)K) \le v+m_T+\lambda_L \le K+T-2$.
This is equivalent to $(L-2)K+1 \le 0$, a contradiction, hence $N > KL+K+T-1$.

The proof of the bound in~\eqref{eq:bound_3} makes use of the fact, cf.~\cite{DegTable/DOliveiraRHK21},
\begin{align*}
|(\alpha_R^{(1)} + \beta_I) \cap (\alpha_I + \beta_R^{(i)})| \le 1
\end{align*}
for all $i=1,\ldots,T$:
Assume the contrary, $u \ne v \in (\alpha_R^{(1)} + \beta_I) \cap (\alpha_I + \beta_R^{(i)})$, then
\begin{align*}
&
u = \alpha_R^{(1)} + \beta_I^{(k)} = \alpha_I^{(\ell)} + \beta_R^{(i)}
\text{ and}
\\&
v = \alpha_R^{(1)} + \beta_I^{(k')} = \alpha_I^{(\ell')} + \beta_R^{(i)}
, \text{ hence}
\\&
u-v = \beta_I^{(k)}-\beta_I^{(k')} = \alpha_I^{(\ell)} -\alpha_I^{(\ell')}
, \text{ so that}
\\&
\alpha_I^{(\ell')}+\beta_I^{(k)} = \alpha_I^{(\ell)}+\beta_I^{(k')}
,
\end{align*}
contradicting the uniqueness of elements in $A$.

Then, for $1 \le m \le T$:
\begin{align*}
N
&
=
KL+|B \cup C|
\\&\ge
KL+|B \cup ((\alpha_R^{(1)},\ldots,\alpha_R^{(m)})+\beta_I)|
\\&=
KL+|B|+|(\alpha_R^{(1)},\ldots,\alpha_R^{(m)})+\beta_I|\\
&-|B \cap ((\alpha_R^{(1)},\ldots,\alpha_R^{(m)})+\beta_I)|
\\&=
KL+(K+T-1)+(L+m-1)\\
&-|B \cap ((\alpha_R^{(1)},\ldots,\alpha_R^{(m)})+\beta_I)|
\end{align*}
and
\begin{align*}
B \cap ((\alpha_R^{(1)},\ldots,\alpha_R^{(m)})+\beta_I)
=
\bigcup_{i=1}^m \left( B \cap (\alpha_R^{(i)}+\beta_I) \right)
\end{align*}
as well as
\begin{align*}
&
B \cap (\alpha_R^{(i)}+\beta_I)
=
(\alpha_I+\beta_R) \cap (\alpha_R^{(i)}+\beta_I)\\
&=
\left(\bigcup_{j=1}^{T}(\alpha_I+\beta_R^{(j)})\right) \cap (\alpha_R^{(i)}+\beta_I)
\\&=
\bigcup_{j=1}^{T}\left((\alpha_I+\beta_R^{(j)}) \cap (\alpha_R^{(i)}+\beta_I)\right)
\end{align*}
so that
\begin{align*}
|B \cap (\alpha_R^{(i)}+\beta_I)|
&=
\left|\bigcup_{j=1}^{T}\left((\alpha_I+\beta_R^{(j)}) \cap (\alpha_R^{(i)}+\beta_I)\right)\right|\\
&\le
\sum_{j=1}^{T}\left|\left((\alpha_I+\beta_R^{(j)}) \cap (\alpha_R^{(i)}+\beta_I)\right)\right|\\
&\le
\sum_{j=1}^{T}1
=
T
.
\end{align*}
Moreover, clearly $|B \cap (\alpha_R^{(i)}+\beta_I)| \le L$.
Hence,
\begin{align*}
|B \cap ((\alpha_R^{(1)},\ldots,\alpha_R^{(m)})+\beta_I)|
&\le
\sum_{i=1}^m \left| B \cap (\alpha_R^{(i)}+\beta_I) \right|\\
&\le
\sum_{i=1}^m \min\{L,T\}\\
&=
m\min\{L,T\}
\end{align*}
and
\begin{align*}
N
\ge&
KL+(K+T-1)+(L+m-1)\\
&-|B \cap ((\alpha_R^{(1)},\ldots,\alpha_R^{(m)})+\beta_I)|
\\\ge&
KL+K+T+L+m-2-m\min\{L,T\}
.
\end{align*}
Exchanging $K$ and $L$, and $\alpha$ and $\beta$ and applying virtually the same argument yields
\begin{align*}
N
\ge
KL+L+T+K+m-2-m\min\{K,T\}
.
\end{align*}
Combining both bounds on $N$ implies now
\begin{align*}
N
\ge&
\max\{KL+K+T+L+m-2-m\min\{L,T\},\\
&KL+L+T+K+m-2-m\min\{K,T\}\}
\\=&
KL+K+L+T+m-2+m\max\{-\min\{L,T\},\\
&-\min\{K,T\}\}
\\=&
KL+K+L+T+m-2-m\min\{\min\{L,T\},\\
&\min\{K,T\}\}
\\=&
KL+K+L+T+m-2-m\min\{K,L,T\}
.
\end{align*}

The last two statements follows from
\begin{align*}
&
KL+K+L+T+m-2-m\min\{K,L,T\} \\
&\le KL+K+L+T-1-\min\{K,L,T\}
\\\Leftrightarrow&
m-2-m\min\{K,L,T\} \le -1-\min\{K,L,T\}
\\\Leftrightarrow&
m-1 \le (m-1)\min\{K,L,T\}
\\\Leftrightarrow&
1 \le \min\{K,L,T\}
\end{align*}
and
\begin{align*}
&
KL+\max\{K,L\}+T-1 \\
&\le KL+K+L+T-1-\min\{K,L,T\}
\\\Leftrightarrow&
\max\{K,L\} \le K+L-\min\{K,L,T\}
\\\Leftrightarrow&
\max\{K,L\} \le \max\{K,L\}+\min\{K,L\}-\min\{K,L,T\}
\\\Leftrightarrow&
\min\{K,L,T\} \le \min\{K,L\}
.
\end{align*}
\end{proof}

\subsection{Proof of Theorem~\ref{thm:OptimalityGap}}\label{pf:OptimalityGap}

\begin{proof}
The optimality gap of $\widehat{N}_{\mathrm{small}}^{\mathrm{pre}}$ minus ``bound in~\eqref{eq:bound_3} with $m=1$'' is
\begin{align}\label{eq:optgap1}
T-3-\left\lfloor\frac{T-2}{\min\{K,L\}}\right\rfloor+\min\{K,L,T\}
\end{align}
and $\widehat{N}_{\mathrm{small}}^{\mathrm{pre}}$ is bound achieving if~\eqref{eq:optgap1} is $\le 0$.
\eqref{eq:optgap1} can be simplified with the fact $x \le \lfloor y \rfloor \Leftrightarrow x \le y$ for any integer $x$ and real number $y$ to
\begin{align*}
\min\{K,L\}(T-3+\min\{K,L,T\}) \le T-2
\end{align*}
and in turn to
\begin{align}\label{eq:optgap2}
\frac{X-1}{X} (T-2) + (x-1) \le 0
\end{align}
by substituting $x=\min\{K,L,T\}$ and $X=\min\{K,L\}$.

If $X=1$ then $x=1$ and~\eqref{eq:optgap2} is true.
Else $X \ge 2$ and hence $T \le 2$.
If $X \ge 2$ and $T = 1$, then $x=1$ and~\eqref{eq:optgap2} is true.
If $X \ge 2$ and $T = 2$, then $x=2$ and~\eqref{eq:optgap2} is false.
Hence, if at least one of $K,L,T$ is 1, then $\widehat{N}_{\mathrm{small}}^{\mathrm{pre}}$ achieves the bound of~\eqref{eq:bound_3} with $m=1$.

Next, we compare $\widehat{N}_{\mathrm{small}}^{\mathrm{pre}}$ to~\eqref{eq:bound_2}, which requires $2 \le \min\{K,L,T\}$.
Calculating again the optimality gap of $\widehat{N}_{\mathrm{small}}^{\mathrm{pre}}$ minus~\eqref{eq:bound_2}, we obtain with the substitution $m=\min\{K,L\}$
\begin{align}\label{eq:optgap3}
m+T-4-\left\lfloor\frac{T-2}{m}\right\rfloor
\end{align}
which shows that $\widehat{N}_{\mathrm{small}}^{\mathrm{pre}}$ is bound achieving if~\eqref{eq:optgap3} is $\le 0$.
Again, \eqref{eq:optgap3} can be simplified with similar transformations as before to
\begin{align*}
T \le \frac{-m^2+4m-2}{m-1}
.
\end{align*}
As $2 \le \min\{K,L,T\}$, we have $2 \le T$ and $2 \le m$, hence $T=2$ (which implies $m=2$) is the only solution to this inequality.

\end{proof}

\subsection{Proof of Theorem~\ref{thm:complexity_no_precomp}}

The proof follows the same technique as presented in~\cite[Theorem~3]{rafael_note}. In Tables~III and IV of~\cite{rafael_note}, the time complexities for both the communication and computation of a polynomial scheme for SDMM are shown. When setting the number of collusions $T$ to be a constant fraction of $N$, the only cost that changes is the encoding cost which becomes $\mathcal{O}(n^{2+3\varepsilon})$, whereas before, when $T$ was constant, it was $\mathcal{O}(n^{2+2\varepsilon})$. Thus, the total time complexity is now the maximum between the encoding complexity and the server multiplication, i.e., $\mathcal{O}(n^{\max\{\varepsilon+\omega-\varepsilon\omega,2+3\varepsilon\}})$.

\subsection{Proof of Theorem~\ref{thm:complexity_with_precomp}}

The proof follows the same technique as presented in~\cite[Theorem~3]{rafael_note}. In Tables~III and IV of~\cite{rafael_note}, the time complexity for both the communication and computation of a polynomial scheme for SDMM are shown. When we precompute the random products, we can precompute the evaluations of $f_R$ and $g_R$ in~\eqref{eq:f_g_polys}. Thus, the computational complexity of the encoding becomes equal to that of Table~IV in~\cite{rafael_note}, but with $T = 1$. Thus, the total time complexity is now the maximum of the encoding time and the server multiplication, i.e., $\mathcal{O}(n^{\max\{\varepsilon+\omega-\varepsilon\omega,2+2\varepsilon\}})$.

\subsection{Proof of Theorem~\ref{thm:grasp4L4}}\label{pf:grasp4L4}

Note that $A + B = [0:4L-1]$, so every desired degree has a unique representation.
Moreover,
\[
    \min(A + S) = \min(R + B) = 4L, \quad \min(R + S) = 8L,
\]
so no other quadrant intersects $A+B$.
The entries of $R$ and $S$ are pairwise distinct, proving decodability and $4$-security over a sufficiently large field extension.

For the threshold count,
\begin{align*}
    R + B &= 4L + 2[0:L+2], \\
    A + S &= [4L: 4L+4] \cup [6L: 6L+4],
\end{align*}
where every even element of $A+S$ already belongs to $R+B$.
The only new elements are
\[
    4L+1, \quad 4L+3, \quad 6L+1, \quad 6L+3.
\]
Hence
\[
    \left|(A + S) \cup (R + B)\right| = L + 7.
\]
This union is disjoint from $A+B$, which has size $4L$, and therefore
\[
    \Npre = 4L + (L + 7) = 5L + 7.
\]

\subsection{Proof of Theorem~\ref{thm:4L4Npre}}\label{pf:4L4Npre}

Let
\[
    C=R+B, \quad E=A+S, \quad q=|C\cup E|.
\]
Since $A+B$ has $4L$ unique degrees and is disjoint from $C\cup E$,
\[
    \Npre=4L+q.
\]
For $z\in\mathbb Z$, define
\begin{align*}
    c_z &= \bigl|\{(r,b) \in R \times B \mid r+b=z\}\bigr|, \\
    e_z &= \bigl|\{(a,s) \in A \times S \mid a+s=z\}\bigr|.
\end{align*}
For each fixed $(r,s)\in R\times S$, there is at most one solution to $r+b=a+s$.
Two distinct solutions would imply a collision in $A+B$.
Thus
\begin{equation}\label{eq:czezSum}
    \sum_z c_z e_z \leq |R| |S| = 16.
\end{equation}
The integer sumset bound gives
\begin{equation}\label{eq:CEBound}
    |C| \geq L+3, \quad |E| \geq 7.
\end{equation}
Assume that $q\leq L+4$.  Let
\[
    I = C \cap E, \quad i=|I|, \quad x = |C \setminus E|, \quad y = |E \setminus C|.
\]
By~\eqref{eq:CEBound}, $i\geq 6$.
Since $c_z,e_z\leq 4$,
\[
    \sum_{z\in I}c_z\geq 4L-4x, \quad \sum_{z\in I}e_z\geq 16-4y.
\]
For positive integers $c_z, e_z$, one has $c_z e_z \geq c_z + e_z - 1$.  Therefore
\[
    \begin{aligned}
        \sum_{z \in I} c_z e_z & \geq 4L + 16 - 4x - 4y - i \\
        &= 4L + 16 - 4q + 3i \\
        &\geq 18,
    \end{aligned}
\]
contradicting~\eqref{eq:czezSum}.
Hence $q \geq L+5$, and so $\Npre \geq 5L+5$.

\subsection{Proof of Lemma~\ref{lem:span_bounds}}\label{pf:span_bounds}

\begin{proof}
    Translate $P$ and $Q$ so that $\min\{P\} = \min\{Q\} = 0$, then set $d = \max\{P \cup Q\}$.
    The $K-1$ positive elements of $P$ and the $L-1$ positive elements of $Q$ are differences in $P-P$ and $Q-Q$ respectively, hence disjoint subsets of $[1:d]$, so $d \geq K + L - 2$.
    Moreover, $P + Q \subseteq [0:2d]$ has $KL$ distinct elements by injectivity, so $KL \leq 2d + 1$.

    For the upper bound, let $a = \lceil L/2 \rceil$ and $b = L - a$, then take the split code of Definition~\ref{def:GRASP} in the orientation $(L,K)$:
    \[
        P = a \cdot [0:K-1], \quad Q = [0:a-1] \cup \bigl(aK + [0:b-1]\bigr).
    \]
    Every nonzero difference of $P$ is a nonzero multiple of $a$, with absolute value at most $a(K-1)$.
    A difference between two elements of the same block of $Q$ has absolute value at most $a-1$, and a difference between the two blocks has absolute value at least $aK - (a-1) = a(K-1) + 1$.
    Thus, $(P-P) \cap (Q-Q) = \{0\}$, and $\max\{P \cup Q\} = aK+b-1 = (K-1) \lceil L/2 \rceil + L - 1$.
    Interchanging $K$ and $L$ provides the other term for $\sigma(K,L)$.
\end{proof}

\subsection{Proof of Theorem~\ref{thm:span}}\label{pf:span}

If $p+q = p'+q'$, then $p-p' = q'-q \in (P-P) \cap (Q-Q)$, so $p=p'$ and $q=q'$, implying that the $KL$ desired degrees $P+Q$ are distinct.
They are elements of $[0:2s]$, with $c > 2s$, so every degree in the other three quadrants exceeds every desired degree.
The random exponents are pairwise distinct in both encoding polynomials, so decodability and $T$-security follow from the degree-table criterion of Section~\ref{sec:gasp-precomputation}.
%
%{sec:GsmallGbig}.
After precomputing the random-random quadrant, the two mixed supports are $c + (P + [0:T-1])$ and $c + (Q + [0:T-1])$, with union $c + \bigl( (P \cup Q) + [0:T-1] \bigr)$.
If $P \cup Q = \{u_0 < u_1 < \dots < u_h\}$, where $u_0 = 0$ and $u_h = s$, then this union has $T + \sum\limits_{i < h} \min\{T, u_{i+1} - u_i\}$ elements.
This is at most $T + s$, with equality when every gap is at most $T$.

\subsection{Proof of Theorem~\ref{thm:span_four}}\label{pf:span_four}

The upper bounds are $\sigma(4,2) = 4$ and the pair $P = \{0, 1, 2L, 2L+1\}$, $Q = 2 \cdot [0:L-1]$ above.
Lemma~\ref{lem:span_bounds} provides the lower bound $s^*(4,L) \geq 2L$.
Suppose $s = 2L$ is achieved by $P,Q \subseteq [0:2L]$, with $0 \in P \cap Q$.
Then $P+Q$ has $4L$ distinct elements in $[0:4L]$, so exactly one nonzero element of $[0:4L]$ is absent -- call it $e$.
If $0 < e < 4L$, then $4L \in P+Q$ forces $p = q = 2L$, so $2L \in (P-P) \cap (Q-Q)$, which is a contradiction.
If $e = 4L$, then $P \oplus Q = [0:4L-1]$, therefore Lemma~\ref{lem:deBruijn} applies with $n = 4L$, $X = P$, and $Y = Q$.
The largest index $t$ contributes $w_t (f_t - 1) = 4L - \tfrac{4L}{f_t} \geq 2L$ to the maximum of whichever set contains it, with equality only if $f_t = 2$.
If $t \in I_P$, then since $|P| = 4$, either $f_t = 4$ and $\max\{P\} = 3L$, or $f_t = 2$ so $I_P$ contains a second index of weight at least $1$, hence $\max\{P\} \geq 2L + 1$.
If $t \in I_Q$, then $f_t$ divides $L$, so either $f_t \geq 3$ and $\max\{Q\} \geq \tfrac{8L}{3} > 2L$, or $f_t = 2$ and the remaining indices of $I_Q$ have product $L/2 \geq 2$, thus $\max\{Q\} \geq 2L + 1$.
No matter what, every case has $s \geq 2L + 1$.

\subsection{Proof of Theorem~\ref{thm:span_vs_gasp}}\label{pf:span_vs_gasp}

    In the orientation with information supports $[0:K-1]$ and $K \cdot [0:L-1]$, Theorem~\ref{thm:GasprServers1} can be written as $\Npre_r = KL + (K+T-1) + [T + (L-1-V)r - \min\{r, K_0\}]_+$, where $K + T - 1 = VK + K_0$ for $0 \leq K_0 < K$.
    For $r<K$, the bracketed term equals $T - rT/K + O_{K,L,r}(1)$, so $\Npre_r$ grows like order $(2 - r/K)T$ with a coefficient larger than one, while $r=K$ gives $2KL - K + T$ for $T \geq K$ by~\eqref{Eq:GbigNpre_symm}.
    There are finitely many values of $r$ and only two orientations, so for large $T$, the best choice is $r=K$ in the orientation with $K=M$, giving $2KL - M + T$.

    Assume $K=m$ and $L=M$.
    The second term of $\sigma(K,L)$ is $(M-1) \lceil m/2 \rceil + m - 1$, and
    \[
        M(m-1) - \Bigl( (M-1) \Bigl\lceil \frac{m}{2} \Bigr\rceil + m - 1 \Bigr) = (M-1) \Bigl( \Bigl\lfloor \frac{m}{2} \Bigr\rfloor - 1 \Bigr),
    \]
    which is positive for $m \geq4$.
    Since $KL - M = M(m-1)$, this gives the claimed expression.
    If $m=3$ and $M$ is even, the first term of $\sigma(K,L)$ is $2 \lceil M/2 \rceil + M - 1 = 2M - 1 < 2M = KL - M$.
    In both cases, $KL + T + \sigma(K,L) < 2KL - M + T$ for all large $T$, and the span construction with the split pair attains the left-hand side by Theorem~\ref{thm:span}.

\end{document}